\documentclass[10pt, twocolumn]{IEEEtran}

\usepackage{xspace,amsmath,amssymb,amsfonts,epsfig,syntonly}
\usepackage{cite,bm,color,url,textcomp,empheq,boxedminipage}
\usepackage{algorithmicx,algorithm,algpseudocode}
\usepackage{epstopdf}
\usepackage{empheq}
\usepackage{pifont}

\usepackage{graphicx,graphics}  
\usepackage{multirow,multicol}
\usepackage{psfrag}    
\usepackage{stfloats}
\usepackage{url}

\newtheorem{theorem}{\underline{Theorem}}[section]
\newtheorem{lemma}{\underline{Lemma}}[section]

\newtheorem{proposition}{\underline{Proposition}}[section]

\long\def\symbolfootnote[#1]#2{\begingroup
\def\thefootnote{\fnsymbol{footnote}}
\footnote[#1]{#2}\endgroup}

\psfull

\allowdisplaybreaks[4]

\usepackage[T1]{fontenc}
\usepackage{ifpdf}
 \ifpdf
 \else
 \fi

\begin{document}

\title{Optimal Pricing of User-Initiated Data-Plan Sharing in A Roaming Market}

\author{Feng Wang, \IEEEmembership{Member,~IEEE,} Lingjie Duan, \IEEEmembership{Senior Member,~IEEE,} and Jianwei Niu, \IEEEmembership{Senior Member,~IEEE} \\

\thanks{Manuscript received December 18, 2017; revised March 2, 2018; accepted June 24, 2018. The work was supported in part by Singapore Ministry of Education Academic Research Fund Tier 2 under Grant MOE2016-T2-1-173, and in part by the National Natural Science Foundation of China under Grants 61572060 and 61772060. This paper was presented in part at the IEEE Global Communications Conference, Singapore, December 4--9, 2017~\cite{Conf17}. The associate editor coordinating the review of this paper and approving it for publication was W. Saad. \emph{(Corresponding author: Lingjie Duan.)}}

\thanks{F. Wang is with the School of Information Engineering, Guangdong University of Technology, Guangzhou 510006, China, and also with the Engineering Systems and Design Pillar, Singapore University of Technology and Design, Singapore 487372 (e-mail: fengwang.nl@gmail.com; fengwang13@gdut.edu.cn).}

\thanks{L. Duan is with the Engineering Systems and Design Pillar, Singapore University of Technology and Design, Singapore 487372 (e-mail: lingjie\_duan@sutd.edu.sg).}

\thanks{J. Niu is with Beijing Advanced Innovation Center for Big Data and Brain Computing (BDBC), the State Key Laboratory of Virtual Reality Technology and Systems, Beihang University, Beijing 100191, China (e-mail: niujianwei@buaa.edu.cn).}}

\maketitle

\begin{abstract}
 A smartphone user's personal hotspot (pH) allows him to share cellular connection to another (e.g., a traveler) in the vicinity, but such sharing consumes the limited data quota in his two-part tariff plan and may lead to overage charge. This paper studies how to motivate such pH-enabled data-plan sharing between local users and travelers in the ever-growing roaming markets, and proposes pricing incentive for a data-plan buyer to reward surrounding pH sellers (if any). The pricing scheme practically takes into account the information uncertainty at the traveler side, including the random mobility and the sharing cost distribution of selfish local users who potentially share their pHs. Though the pricing optimization problem is non-convex, we show that there always exists a unique optimal price to tradeoff between the successful sharing opportunity and the sharing price. We further generalize the optimal pricing to the case of heterogeneous selling pHs who have diverse data usage behaviors in the sharing cost distributions, and we show such diversity may or may not benefit the traveler. Lacking selfish pHs' information, the traveler's expected cost is higher than that under the complete information, but the gap diminishes as the pHs' spatial density increases. Finally, we analyze the challenging scenario that multiple travelers overlap for demanding data-plan sharing, by resorting to a near-optimal pricing scheme. We show that a traveler suffers as the travelers' spatial density increases.
\end{abstract}



\section{Introduction}

\subsection{Background and Motivation}
Roaming is the ability of customers to use their mobile devices outside the geographical coverage area provided by their normal network operator \cite{GSMA}. As the penetration of smart mobile devices increases fast, the volume of the global data roaming market has grown by more than sixfold in the past five years\cite{GSMA,ITU,Tsunami,Cisco}. Accordingly, the revenue of data roaming is expected to increase to US\$50 billions in revenues by 2019 (see \cite{Informa,Cisco}). Travelers may suffer a ``bill shock'', and data roaming is typically expensive as compared to the domestic markets' two-part tariff data plans for local users. To reduce the roaming cost, a traveler may enjoy data services by accessing to the public (free) WiFi hotspots. However, free WiFi hotspots are limited in coverage and they usually concentrate on the public service areas (e.g., airports, stations, malls, and public libraries) to avoid formidably high deployment cost for full coverage. For example, even for a populous city such as Singapore, the WiFi coverage percentage is only 35\% \cite{FreeWiFi}. It is urgent to find economically viable approaches for providing travelers with ubiquitous wireless data services.

With the recent techonology advancements, cellular-enabled iPhones and Android phones can now set up personal hotspots (pHs) to share there cellular data connections with nearby wireless devices (e.g., phones, laptops, and tablets) \cite{PH}. The physical coverage of a pH ranges about hundreds of feet like WiFi and is expected to keep increasing. However, the development of this user-initiated data-plan sharing still lacks a clear business model and a local selfish user is only willing to share the pH connection with his own devices\cite{wang2016user}. As such, it is important to propose incentive schemes for pH-enabled data-plan sharing in a broad popularity.

\begin{figure}
  \centering
  \includegraphics[width = 3.5in]{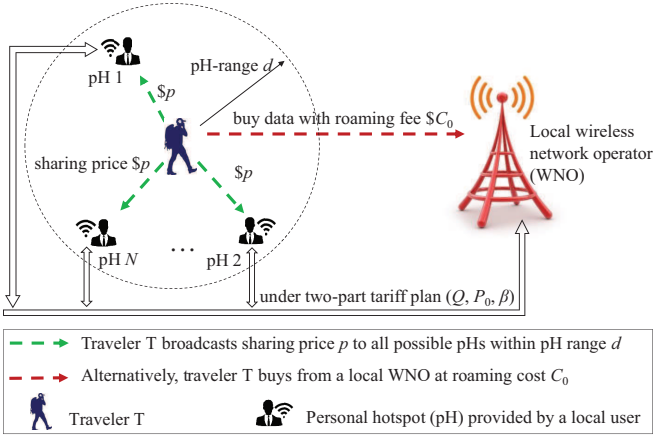}
 \caption{A user-initiated data-plan sharing model with one traveler and personal hotspots (pHs).} \label{fig.Sys-model}
 \end{figure}

A key problem to hinder the development of such data-plan sharing is the sharing cost. A selling user's sharing consumes his monthly data quota in his two-part tariff plan and may lead to overage charge. After subscribing to a data-plan denoted by $(Q,P_0,\beta)$, a local user is given a monthly data quota $Q$ at a fixed lump-sum fee $P_0$ and should pay for overage data beyond $Q$ at a costly unit price $\beta$. In addition, such pH sharing consumes the local user's finite energy in battery storage and also requires the selling user to stay during the sharing period, incurring another {\em waiting} cost. To facilitate user-initiated data-plan sharing via pH, we aim to design a reward-based pricing scheme for a traveler to fairly cover the pH sharing cost. As shown in Fig.~\ref{fig.Sys-model}, a traveler T opportunistically demands data connection from nearby pHs (if any), by announcing a sharing price \$$p$ (ex ante) as a reward via short-range communications. The nearby pHs then respond to accept or not based on their private costs. Finally, the traveler T selects one pH for cellular connection and pays reward \$$p$ to the activated pH.\footnote{The terms ``selling user'' and ``pH'' are interchangeable in this paper.}

However, this optimal pricing design under information uncertainty is challenging at the traveler side. Since the potential pHs are generally moving, the traveler does not know the exact number of pHs in the sharing area. Furthermore, how to estimate the private costs of pHs is another problem for the traveler. Only a pH knows his own sharing cost, since that it depends on the realized data usage in his two-part tariff plan. In general, different pHs have different data usage behaviors or cost distributions. The traveler would prefer to activate a pH with low cost distribution, yet this screening is \emph{not} doable for ex ante pricing. Intuitively, one can expect that a high sharing payment will provide a good chance for pH sharing but leave a low sharing benefit to the traveler. This motivates our study in this work.

\subsection{Contributions}

In this paper, we first develop a model for pH-enabled data-plan sharing initiated by on-demand travelers and then find the optimal pricing in different scenarios. Our main contributions are summarized as follows.

\begin{itemize}
\item {\em Pricing incentive for pH sharing under information uncertainty}: In Section II, to provide an incentive of data-plan sharing to nearby pHs, we design a pricing scheme for an on-demand traveler to reward pHs and reach a win-win situation. The pricing scheme practically considers the traveler's information uncertainty about pHs, including the mobile pHs' random locations following a Poisson point process (PPP) and their sharing cost distribution under the two-part monthly tariff plans.

\item {\em Benchmark case under complete information}: In Section III, to provide a performance bound and evaluate the proposed pH pricing schemes later, we investigate the {\rm social optimum} benchmark case under complete information, where the pHs nearby are willing to report their locations and private costs. Then the traveler pays and activates the pH with the minimal cost. It shows that the traveler's expected cost decreases in both the pHs' spatial density and their residual data quota.

\item{\em Optimal pricing analysis for various pHs under information uncertainty}: In Section IV, the optimal price is determined for the traveler to minimize his expected cost after sharing. We first consider homogeneous pHs whose statistics of monthly data usage are identical. The globally optimal sharing price is derived for the on-demand traveler. We show that the traveler's final expected cost is larger as compared with the case under complete information and the gap diminishes with the pH spatial density. Furthermore, we extend the optimal pricing scheme to the case of heterogeneous pHs who have different data usage statistics. The optimal price is derived by comparing their diverse cost distributions and targeting for the dominant pH type. This diversity may or may not benefit the traveler.

\item{\em Optimal pricing for overlapping travelers}: In Section V, we further consider the scenario that a traveler may overlap with other travelers in demanding common pHs in the considered area. To handle the non-tractable expression of the expected cost for travelers, a lower bound of the expected cost is pursued to obtain the near-optimal price for coordinating all travelers. We show that each traveler suffers from the increase of the travelers' spatial density.
\end{itemize}

\subsection{Related Work}
It is noted that pricing incentive design has received increasingly attentions for wireless networks \cite{duan2014motivating,sen2013survey}. The works in \cite{Luo17,PanMiao_17,PanMiao_JSAC} surveyed some generic pricing and auctions schemes for mobile crowdsourcing. Some recent works investigated the pricing incentive in data trading and user cooperation in the literature. For example, following game theory frameworks, \cite{mus2006wifi} and \cite{duan2015pricing} studied the optimal pricing for static WiFi hotspots (e.g., in a cafe) according to customer types and network capacity. About mobile users' data plan sharing, \cite{Yu15} and \cite{zheng2015secondary} studied mobile data trading among users under the central coordination of the wireless operator in the operator-controlled secondary market. \cite{guo2016optimal} further investigated the pricing incentive to stimulate users' relay cooperation for energy saving purpose. \cite{Gao14} studied a network-controlled user-provided connectivity system and derived the optimal hybrid pricing-reimbursing policy to maximize the network revenue. \cite{Neely11} considered utility maximization for peer-to-peer networks based on a tit-for-tat incentive mechanism. Building on indirect reciprocity game framework, \cite{Chen11} investigated cooperation stimulation for cognitive networks. \cite{Ai09} proposed a credit based mechanism for WiFi sharing community networks. Leveraging the density and heterogeneity of wireless devices, \cite{Lorenzo17} proposed a new cognitive dynamic architecture for future wireless networks to provide ubiquitous Internet connectivity and developed a distributed matching algorithm for operators' and users' cooperation for data sharing. There are other pricing schemes for non-tethering wireless network \cite{Li16,Rose14}, where the time-dependent pricing for price-quality tradeoff is analyzed.

Different from these operator-controlled data sharing networks, we investigate a user-initiated peer-to-peer network for data trading via pHs without operators' intervention. Note that \cite{wang2016user} investigated a user-initiated data-plan sharing scenario by employing pHs, where users with diverse data usage trade data plans and the wireless operator indirectly intervene with such sharing to control overage charge. However, \cite{wang2016user} assumed complete information for mobile data trading and did not take into account user mobility and private cost in practice. Unlike prior works, this paper studies an optimal pH pricing problem under information uncertainty, where the traveler only has partial information about pHs' private locations and costs.

\section{System Model}\label{Sec:System}
\begin{figure*}
  \centering
  \includegraphics[width = 6.0in]{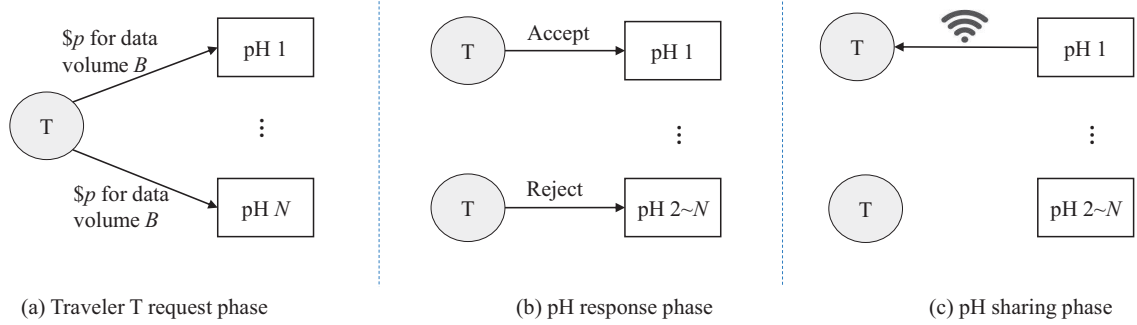}
 \caption{An illustrative example of pH-enabled data-plan sharing between traveler T and $N$ pHs ((if any).} \label{fig.protocol}
\end{figure*}

Consider a typical traveler T who wants to connect the Internet. He can potentially buy data from any neighboring pH within pH range (e.g., around 50 meters). The pH range $d$ is determined by traveler T's quality of service (QoS) requirement on the minimum received signal-to-interference-plus-noise ratio (SINR) target $\gamma>0$. One can equivalently consider the minimum data rate. Note that the traveler's application is QoS-guaranteed type (e.g., video call), and he perceives no service difference one beyond the SINR target. Suppose traveler T is located at the origin. Let $\Lambda(\lambda)=\{\bm y\}$, $\bm y\in \mathbb{R}^2$, denote the coordinates of the pHs. Due to a collection of pHs distributed in the space by the PPP $\Lambda(\lambda)$ and they share a common radio medium, the total power received at traveler T from this collection of pHs is in essence a shot-noise field at the origin. For ease of exposition, we consider a typical signal propagation model accounting for large-scale path loss, and all pHs employ a constant and identical power of $P_{\rm tx}$ without power control. The SINR of traveler T at distance $d$ from its target pH is expressed as \cite{BookPPP,Jeff11}
\begin{align}\label{eq.I_o_value}
{\rm SINR}(d) = \frac{P_{\rm tx} C_0({d}/{r_0} )^{-\alpha}}{I_d+\sigma^2},
\end{align}
where $C_0$ is a constant path loss value at reference distance $r_0$, $\alpha>2$ is the path loss exponent depending on the antenna height and the signal propagation environment \cite{GoldsmithBook}, and $\sigma^2$ is the power of the additive noise at traveler T's receiver. In \eqref{eq.I_o_value}, the term $I_d\triangleq \sum_{{\bm y}\in \Lambda(\lambda)\setminus \{{\bm b}_o\}} P_{\rm tx} C_0({\|\bm y\|}/{r_0} )^{-\alpha}$ is the cumulative interference from all the other pHs (except the target pH for traveler T at location $\bm b_0$) at distance $\|\bm y\|$ from traveler T. By using the null probability of a PPP, the probability density function of $\|\bm y\|$ is then obtained as $f_{\|\bm y\|}(r)=2\pi\lambda r\exp(-\lambda\pi r^2)$ for $r\geq 0$. The Laplace transform of the interference $I_d$ is  \cite{Jeff11}
\begin{align}\label{eq.Id}
&{\cal L}_{I_d}(s) = \notag\\
& \exp\Big(-2\pi\lambda\int_{d}^\infty (1-\exp(-s P_{\rm tx} C_0(r/r_0)^{-\alpha}))rdr\Big).
\end{align}
Based on \eqref{eq.Id}, one can evaluate the interference value $I_d$ for any pH range $d$.

To ensure ${\rm SINR}(d) \geq \gamma$, where $\gamma$ is the minimum SINR target, the maximum value of pH range $d$ is given by
\begin{align}\label{eq.d}
d=\Bigg(\frac{P_{\rm tx}C_0r_0^\alpha}{\gamma(I_d+\sigma^2)}\Bigg)^{\frac{1}{\alpha}}.
\end{align}
From \eqref{eq.d}, under a certain SINR target $\gamma$, it is expected that a larger pH density imposes a stronger interference on traveler T, thereby reducing the pH radius $d$ of the sharing area. Note that similar results of \eqref{eq.I_o_value}--\eqref{eq.d} can be obtained if a more general fading channel is considered, where the received SINR and maximum pH-range can be modified accordingly with a given maximum outage probability and fading distribution \cite{GoldsmithBook,Jeff11}. Note that once the average pH service quality is comparable to the roaming of the cellular network, the traveler will still choose the former service to avoid the high charge in the latter roaming.

At the center of circle area $A$ with radius $d$, traveler T can access to the data service either by paying one activated pH within the area $A$ or by paying the local wireless network operator (WNO). Denote by ${\cal H}$ the set of randomly appeared pH number in the area $A$ and $N\triangleq |\cal H|$.


Let $B$ be the volume of traveler T's requested data to buy. For traveler T, we denote $p$ and $C_0$ as his announced price to pHs nearby and the roaming fee charged by the local WNO, respectively. A selling pH will serve the demanding traveler T within range $d$ and the traveler T will try to activate and connect to a pH within $d$. The pH-enabled data-plan sharing procedure is illustrated in Fig.~\ref{fig.protocol}, and can be divided into three phases:
\begin{itemize}
\item {\em Traveler request phase:} at the beginning, traveler T broadcasts the data-plan sharing request message via short range communications (e.g., bluetooth), where the message contains the offered price $p$ for the data volume $B$.

\item {\em pH response phase:} upon receiving the request, in this phase each pH $i\in {\cal H}$ (if any) makes a decision to accept or not and responds to traveler T.

\item {\em pH sharing phase:} in this phase, if at least one positive pH response message is received, then traveler T will randomly select one of the positive pHs and pay $p$ to establish the pH connection; otherwise, traveler T will resort to the WNO by paying roaming fee $C_0$ for data volume $B$.
\end{itemize}

Note that we pursue pH data-sharing pricing optimization with a fixed-volume data usage for traveler T in this paper. This is reasonable for many inelastic applications, such as file transfer \cite{mus2006wifi}. Once the pH data connection is established between the pH and the traveler, it requires that both the traveler and the pH commit to stay within the distance of $d$ until the data sharing service of a certain data amount (e.g., $B$) is completed. Thus, network change does not affect the existing service linkage. On the other hand, the pHs are moving in general and the private information of pHs (e.g., the number $N$ and data-plan sharing costs) is not available for traveler T {\em ex ante} in traveler request phase.

\subsection{pHs' models about mobility, cost and utility}
To capture the nature of pH mobility, we assume that the locations of pHs in target circle area $A$ follow a two-dimensional PPP with spatial density $\lambda$ as in \cite{Jeff10}. The average number $N=|\cal H|$ of pHs in area $A$ is $\lambda \pi d^2$, which increases in the pH range $d$ and the pH density $\lambda$. The probability mass function (PMF) of $N$ is then
 \begin{equation}\label{eq.HU_number}
 {\rm Pr}(N=n) = \frac{(\lambda\pi d^2)^n}{n!}\exp(-\lambda\pi d^2),~~n=0,1,\cdots
 \end{equation}
 where ${\rm Pr}(X)$ denotes the probability of event $X$. Ideally, traveler T wants to attract only one pH at minimum price but may fail to attract none, depending on the cost distribution of various pHs in sharing.

The data-plan sharing to traveler T will consume pH $i$'s cellular data amount $B$ for any $i\in{\cal H}$. Suppose that pH $i$ has subscribed to an identical two-part tariff plan $(Q,P_{0},\beta)$ from the local WNO in the long run, where $Q$ is the monthly data quota, $P_0$ is the fixed lump-sum fee, and $\beta$ is the unit price for overage data beyond $Q$. In practice, when each pH $i\in{\cal H}$ predicts his monthly data usage $x_i$, he will inevitably attach an additive noise $\Delta_i\sim {\cal N}(0,\delta^2)$ according to his subjective estimation, where $\delta^2$ represents the estimation error variance. For any pH $i\in{\cal H}$, the monthly data usage $x_i$ is assumed to be a Gaussian variable, i.e., $x_i\sim{\cal N}(\mu,\sigma^2)$, where $\mu$ and $\sigma^2$ denote the mean and variance, respectively. By considering the shared volume $B$ with traveler T, the actual monthly usage is $x_i+\Delta_i+B$ for any pH $i\in{\cal H}$, which may be larger than quota $Q$ and hence incurs surcharge with rate $\beta$. Given $x_i$ estimation, the (additional) cost for pH $i$, $i\in{\cal H}$, after sharing with traveler T is
 \begin{align}\label{eq.cdf_Ci}
 C_i(x_i) &={\mathbb E}\{ \beta(x_i+\Delta_i+B-Q)^+ \}- {\mathbb E}\{ \beta(x_i+\Delta_i-Q)^+\} \notag \\
&= \begin{cases}
 0,&~{\rm if}~0\leq x_i \leq Q-B \\
 \beta(x_i+B-Q),&~{\rm if}~Q-B< x_i < Q \\
 \beta B,&~{\rm if}~ x_i\geq Q,
 \end{cases}
 \end{align}
 where $(x)^+\triangleq \max\{x,0\}$ and the expectation ${\mathbb E}\{\cdot\}$ is taken over the estimation noise $\Delta_i$.

Define $\phi(x)$ and $\Phi(x)$ as the probability density function (PDF) and the cumulative distribution function (CDF) of Gaussian variable $x$, respectively. According to (\ref{eq.cdf_Ci}), the minimal cost of pH $i\in{\cal H}$ for data-plan sharing is zero if his monthly data usage $x_i$ plus the sharing data amount $B$ is within the monthly mean data quota $Q$, while the maximal cost is $\beta B$ if his monthly data usage $x_i$ is already larger than the data quota $Q$. Based on \eqref{eq.cdf_Ci} for random $x_i$, we obtain the CDF of the sharing cost $C_i(x_i)\leq \beta B$ in the following lemma:

 \begin{lemma}\label{lem.CDF}
 For pH $i\in{\cal H}$, the CDF of his additional cost $C_i(x_i)$ incurred by traveler T's data consumption $B$ is given by
 \begin{align}
 {\rm Pr}(C_i(x_i)\leq c) =
 \begin{cases}
 \Phi(\frac{c}{\beta}+Q-B),& {\rm if}~ 0\leq c < \beta B\\
 1, &{\rm if}~ c=\beta B.
 \end{cases}
 \end{align}
 \end{lemma}
 \begin{IEEEproof}
 See Appendix A.
 \end{IEEEproof}

 As a return, pH $i$ will receive a reward price $p$ from traveler T. The {\em utility} of pH $i\in{\cal H}$ via data-plan sharing is given by
 \begin{align}
  U_i(x_i) = p-C_i(x_i).
 \end{align}
 Reasonably, we assume that any pH $i\in{\cal H}$ has a {\em reservation utility} $\epsilon>0$ that must be guaranteed to cover both the battery consumption\footnote{How to quantify the transmission energy cost can be found in \cite{Feeny01}, where the energy consumption depends on the transmission time and the volume of transmission data.} and the cost due to waiting time during the connection between the pH and traveler T. In other words, a pH $i$'s total cost for data sharing includes both the expected data sharing cost $C_i(x_i)$ (may incur data overage) and the energy consumption cost, as well as the waiting cost. Hence, pH $i$ accepts traveler T's sharing request only if $U_i(x_i)\geq\epsilon$.

 The traveler has to buy data from the WNO at cost $C_0$ if the following two conditions hold:
 \begin{itemize}
 \item No pH exists in area $A$, i.e., $N=0$;
 \item The utility of each pH $i\in{\cal H}$ in the non-empty set ${\cal H}$ is smaller than $\epsilon$, i.e., $p-C_i(x_i)< \epsilon$.
 \end{itemize}
The realized cost for traveler T is given as
 \begin{equation}\label{eq.Traveler_cost}
 C_{\rm T} =
 \begin{cases}
 p, &~~{\rm if}~ \exists i\in{\cal H},~p-C_i(x_i) \geq \epsilon\\
 C_0,&~~{\rm otherwise},
 \end{cases}
 \end{equation}

To ensure the mutual benefits of traveler T and pHs in this user-initiated data-plan sharing, the offer price $p$ by the traveler T should satisfy
 \begin{equation} \label{eq.box1_p}
 \epsilon \leq p \leq C_0,
 \end{equation}
 where the first inequality in \eqref{eq.box1_p} guarantees the utility to increase of the pHs and the second inequality in \eqref{eq.box1_p} saves the roaming cost for traveler T.

\section{Benchmark Case under Complete Information}\label{sec:Benchmark}
In this section, we consider the ideal benchmark case under complete information, where the set $\cal H$ and the private information of pH $i\in{\cal H}$ (including the expected monthly data usage $x_i$ and the expected cost $C_i(x_i)$ in sharing) is available for traveler T. This case can happen when both traveler T and the pHs belong to a cooperative community (e.g., family members or friends) or pHs are altruistic by reporting their locations and private costs accurately. This case serves as a performance benchmark (lower bound for the traveler's offered price), when comparing to Sections IV and V under incomplete information.

 Under complete information, if pH set $\cal H$ is non-empty, traveler T only offers a price to cover the reservation utility $\epsilon$ plus the minimum cost among all pHs, i.e.,
 \begin{align}\label{eq.p_min}
 p(N) = \epsilon+\min_{i\in{\cal H}} C_i(x_i).
 \end{align}
 Without loss of generality, we reorder the $N$ pHs' costs according to $C_1\leq C_2\leq \ldots\leq C_N$, where $C_i$ is the $i$th smallest pH sharing cost and depends on $x_i$ realization. Then \eqref{eq.p_min} reduces to
 \begin{align}\label{eq.p_complete}
 p(N) = \epsilon+C_1.
 \end{align}
 Based on Lemma~\ref{lem.CDF}, it follows that the CDF of $C_1$ is \cite{book_order_statistic}
 \begin{align}\label{eq.cdf_c1}
 &{\rm Pr}(C_1\leq c) = \notag \\
 &\quad\quad \begin{cases}
  1-\big(1-\Phi(\frac{c}{\beta}+Q-B)\big)^N,& {\rm if}~0\leq c<\beta B\\
  \left(1-\Phi(Q)\right)^N,&{\rm if}~c=\beta B.
 \end{cases}
 \end{align}
and the PDF of $C_1$ is
\begin{align}\label{eq.pdf_c1}
f_{C_1}(c) &=\frac{\partial}{\partial c}{\rm Pr}(C_1\leq c) \notag \\
 &=\frac{N}{\beta}\big(1-\Phi(\frac{c}{\beta}+Q-B)\big)^{N-1}\Phi(\frac{c}{\beta}+Q-B)
\end{align}
for $0\leq c<\beta B$.

Based on \eqref{eq.cdf_c1} and \eqref{eq.pdf_c1}, we have the following lemma.
\begin{lemma}\label{lem.c1}
Given the non-empty pH set (i.e., $N\geq 1$), the expected cost $p(N)$ of traveler T is given by
\begin{align}\label{eq.lem1}
p(N)&=\mathbb{E}\{\epsilon+C_1\} \notag \\
&= \epsilon+\int_{0}^{\beta B}\Big(1-\Phi(\frac{c}{\beta}+Q-B)\Big)^N dc,
\end{align}
where the expectation is taken over any possible $C_1$ with the PDF in \eqref{eq.pdf_c1}.
\end{lemma}
\begin{IEEEproof}
See Appendix B.
\end{IEEEproof}
Lemma \ref{lem.c1} shows that the traveler T's cost $p(N)$ is a decreasing function with respect to $N\geq 1$ due to the fact that $\Phi(\frac{c}{\beta}+Q-B)\in[0,1]$ for $0\leq c \leq \beta B$. Further, denoted by $EC^{**}$ the expected cost of traveler T over any possible pH number $N$ (including $N=0$). Taking expectation of \eqref{eq.lem1} over random $N$, i.e., $EC^{**}=C_0\times {\rm Pr}(N=0)+\sum_{n=1}^\infty p(N)\times {\rm Pr}(N=n)$, we establish the following theorem with the use of iterated expectation \cite{Fishburn70}.
\begin{theorem}\label{theo.c1}
The expected cost $EC^{**}$ of traveler T under complete information is given by
\begin{align}\label{eq.EC_star}
EC^{**}
&= \epsilon +(C_0-\epsilon-\beta B)\exp(-\lambda\pi d^2) \notag \\
&+ \int_{0}^{\beta B} \exp\Bigg(\frac{\lambda\pi d^2\Big({\rm erfc}\Big(\frac{x/\beta+Q-B-\mu}{\sqrt{2}\sigma}\Big)-2\Big)}{2}\Bigg)dx,
\end{align}
where ${\rm erfc}(x)\triangleq \frac{2}{\sqrt{\pi}}\int_{x}^{+\infty}e^{-t^2}dt$ is a complementary error function.
\end{theorem}
\begin{IEEEproof}
See Appendix C.
\end{IEEEproof}


Due to the fact that the function ${\rm erfc}(x)$ decreases with $x$ and by checking the relationship between $EC^{**}$ and some key parameters in Theorem \ref{theo.c1}, we have the following proposition.
\begin{proposition}\label{prop1}
The traveler T's benchmark (minimum) cost $EC^{**}$ always increases with the sharing data amount $B$, the roaming fee $C_0$, and the overage charge unit price $\beta$ for the pHs. Meanwhile, $EC^{**}$ decreases with the pH density $\lambda$ and the pH leftover data amount $Q-\mu$.
\end{proposition}

If $B+\mu \leq Q$, as $\sigma^2$ increases, $\exp\big(\frac{\lambda\pi d^2\big({\rm erfc}\big(\frac{x/\beta+Q-B-\mu}{\sqrt{2}\sigma}\big)-2\big)}{2}\big)$ in \eqref{eq.EC_star} increases, i.e., pH's mean usage $\mu$ plus the sharing data $B$ is still within the data quota $Q$. This indicates that the pH is more likely to incur data overage with an increasing variance, and traveler T has to increase its price or turn to paying roaming fee to the WNO. On the other hand, if $B+\mu > Q$, traveler T may or may not benefit from the increasing $\sigma^2$.

\section{Traveler's Optimal Pricing under Information Uncertainty}\label{sec:Monopoly}
In this section, we aim to optimize the traveler T's pricing under pH information uncertainty to minimize his expected cost, which depends on pHs' locations and cost distributions. As in Section \ref{sec:Benchmark}, we focus on a typical traveler T and assume he has no overlap demand with another traveler. This is reasonable for many places less traveled, and will be generalized in Section~V to include overlapped travelers. We first consider {\em homogeneous} pHs' monthly data usage to analyze the optimal price, and then extend to the more challenging case of heterogeneous pHs' usage behavior.

\subsection{Pricing towards pHs of homogeneous data usage}
Consider a homogeneous pH case, where the monthly data usage $x_i\sim{\cal N}(\mu,\sigma^2)$ of pH $i\in{\cal H}$ is i.i.d. Gaussian distributed with identical mean usage and variance among all pHs. To derive the traveler's expected cost, we first analyze the pHs' acceptance or not on the traveler's request.

\subsubsection{The successful probability for motivating pH data sharing}

 \begin{figure*}
 \centering
 \includegraphics[width = 6.5in]{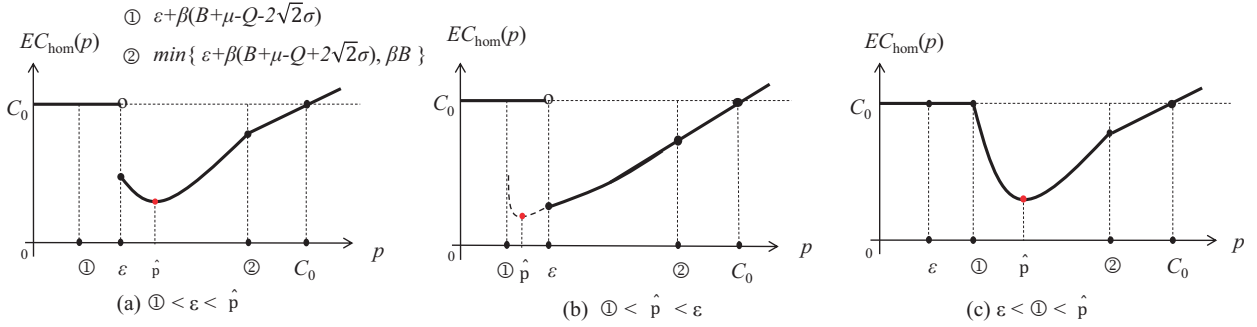}
 \caption{Illustrations of function $EC_{\rm hom}(p)$ for the homogeneous pHs.} \label{fig.curve1}
 \end{figure*}

Denote by $\mathbb{P}_{\tt hom}(p|N)$ the probability for traveler T to successfully attract a pH for data-plan sharing given the pH set $\cal H$. We then have
\begin{align}\label{eq.p_hom}
\mathbb{P}_{\tt hom}(p|N) = 1-\big(1-{\rm Pr}(p-C_i(x_i)\geq \epsilon)\big)^N,
\end{align}
where ${\rm Pr}(p-C_i(x_i) \geq \epsilon)$ is the probability that each pH $i\in{\cal H}$ accepts the data sharing request from traveler T. Note that $\mathbb{P}_{\tt hom}(p|N)$ in \eqref{eq.p_hom} equals zero if $N=0$. Based on the traveler T's estimated PDF of pH $i$'s sharing cost $C_i(x_i)$ in (\ref{eq.cdf_Ci}), the data-plan sharing probability $\tilde{\Omega}(p)$ of each pH is expressed as
\begin{align}\label{eq.AccProb}
\tilde{\Omega}(p) &\triangleq {\rm Pr}(p-C_i(x_i) \geq \epsilon)\notag\\
&= \begin{cases}
0, &{\rm if}~p < \epsilon\\
\Omega(p),&{\rm if}~\epsilon \leq p \leq \epsilon + \beta B\\
1,&{\rm if}~p > \epsilon + \beta B.\\
\end{cases}
\end{align}
where
\begin{align}\label{eq.prob_A}
\Omega(p) &\triangleq  1-\frac{1}{2}{\rm erfc}\big( \frac{p-\epsilon+\beta(Q-B-\mu)}{\sqrt{2}\sigma\beta}
\big)\notag\\
&\begin{cases}
=0,&{\rm if}~p-\epsilon \leq \beta(B+\mu-Q-2\sqrt{2}\sigma)\\
\in(0,1),&{\rm if}~p-\epsilon \in \Big(\beta(B+\mu-Q-2\sqrt{2}\sigma), \\
&\quad\quad\quad\quad\quad \beta(B+\mu-Q+2\sqrt{2}\sigma)\Big)\\
=1,&{\rm if}~p-\epsilon \geq \beta(B+\mu-Q+2\sqrt{2}\sigma).
\end{cases}
\end{align}
Base on \eqref{eq.AccProb} and \eqref{eq.prob_A}, we obtain $\tilde{\Omega}(p)=$
\begin{align}\label{eq.AccProb_new}
 \begin{cases}
0,&{\rm if}~p-\epsilon < (\beta(B+\mu-Q-2\sqrt{2}\sigma))^+\\
\Omega(p), &{\rm if}~p-\epsilon \geq(\beta(B+\mu-Q-2\sqrt{2}\sigma))^+\\
& {\rm and}~p-\epsilon \leq \min\{\beta(B+\mu-Q+2\sqrt{2}\sigma),\beta B\}\\
1,&{\rm if}~p-\epsilon \geq \min\{\beta(B+\mu-Q+2\sqrt{2}\sigma),\beta B\}.\\
\end{cases}
\end{align}
From (\ref{eq.AccProb_new}), for any pH $i\in{\cal H}$, we have the following cases:
\begin{itemize}
\item If $p < \epsilon+(\beta(B+\mu-Q-2\sqrt{2}\sigma))^+$, i.e., the sharing price $p$ is smaller than the given reservation utility plus the minimal possible overage cost, the pH $i\in{\cal H}$ will definitely reject the data sharing request.
\item If $\epsilon+(\beta(B+\mu-Q-2\sqrt{2}\sigma))^+ \leq p \leq \epsilon+\min\{\beta(B+\mu-Q+2\sqrt{2}\sigma),\beta B\}$, the probability in \eqref{eq.prob_A} shows that one pH becomes positive in sharing data connection. This sharing probability increases in the price $p$ and his expected leftover data $Q-B-\mu$, but decreases in its unit price $\beta$ for overage data.
\item If $p > \epsilon+\min\{\beta(B+\mu-Q+2\sqrt{2}\sigma),\beta B\}$, i.e., the sharing price of traveler T is large enough to cover the maximal possible overage cost $\epsilon+\min\{\beta(B+\mu-Q+2\sqrt{2}\sigma), \beta B\}$, any pH $i\in{\cal H}$ will accept the data sharing request and sell data to traveler T.
\end{itemize}
Considering all the possibilities of $N\geq 0$, let $\mathbb{P}_{\tt hom}(p)$ be the expected probability for traveler T to successfully establish the data connection via pH, which is a function of price $p$ offered by traveler T. By taking the expectation of $\mathbb{P}_{\tt hom}(p|N)$ over any possible variable $N$, we have
\begin{align}\label{eq.p_star}
\mathbb{P}_{\tt hom}(p) &=\sum_{N=0}^{\infty}\left(1-(\tilde{\Omega}(p))^N\right) \frac{(\lambda\pi d^2)^N\exp(-\lambda\pi d^2)}{N!}\notag \\
&=1-\exp\left(-\lambda\pi d^2\tilde{\Omega}(p)\right).
\end{align}
By substituting \eqref{eq.AccProb_new} into \eqref{eq.p_star}, it follows that $\mathbb{P}_{\tt hom}(p) = $
\begin{align}
\begin{cases}
 0,\quad &{\rm if}~ p-\epsilon < (\beta(B+\mu-Q-2\sqrt{2}\sigma))^+\\
1-e^{-\lambda \pi d^2\Omega(p)}, &{\rm if}~p-\epsilon \geq(\beta(B+\mu-Q-2\sqrt{2}\sigma))^+\\
&{\rm and}~ p-\epsilon \leq \min\big( \beta B,\\ &\quad\quad\quad\quad\quad \beta(B+\mu-Q+2\sqrt{2}\sigma) \big)\\
1-e^{-\lambda\pi d^2},&{\rm if}~p-\epsilon \geq \min\big(\beta B,\\
& \quad\quad\quad\quad\quad \beta(B+\mu-Q+2\sqrt{2}\sigma) \big),
\end{cases}
\end{align}
which also has three cases as \eqref{eq.AccProb} and in general increases with $p$.

\subsubsection{Traveler T's pricing problem}

 Denote by $EC_{\rm hom}(p)$ the expected cost function of price $p$ for traveler T in the homogeneous pH sharing service. Depending on the response of pHs (if any), the traveler T has either realized cost $p$ or roaming fee $C_0$ in \eqref{eq.box1_p}. Thus, it follows that
\begin{align}\label{eq.expect_cost_traveler}
& EC_{\rm hom}(p) = p\times \mathbb{P}_{\tt hom}(p) + C_0\times(1-\mathbb{P}_{\tt hom}(p)) \notag \\
&=\begin{cases}
  C_0,&{\rm if}~ p-\epsilon < (\beta(B+\mu-Q-2\sqrt{2}\sigma))^+ \\
 \widetilde{EC}_{\rm hom}(p),&{\rm if}~p-\epsilon \geq(\beta(B+\mu-Q-2\sqrt{2}\sigma))^+\\
& {\rm and}~p-\epsilon \leq \min (\beta B,\\
& \quad\quad\quad\quad \beta(B+\mu-Q+2\sqrt{2}\sigma) )\\
 C_0+(p-C_0)&\\
 \times(1-e^{-\lambda\pi d^2}), &{\rm if}~p-\epsilon \geq \min\big(\beta B,\\
 &\quad\quad\quad\quad \beta(B+\mu-Q+2\sqrt{2}\sigma)\big).
\end{cases}
\end{align}
where $\widetilde{EC}_{\rm hom}(p) \triangleq C_0+ (p-C_0)\left(
 1-e^{-\lambda\pi d^2\Omega(p)}\right)$. By checking the first-order and second-order derivatives of $\widetilde{EC}_{\rm hom}(p)$, it is verified that $\widetilde{EC}_{\rm hom}(p)$ is a convex function with respect to $p\in[\beta(B+\mu-Q-2\sqrt{2}\sigma),\min\{\beta(B+\mu-Q+2\sqrt{2}\sigma),\beta B\}]$ (see Appendix D).
In the following, we are interested in minimizing traveler T's expected cost $EC_{\rm hom}(p)$ for $\epsilon\leq p \leq C_0$. Mathematically, we formulate the following optimization problem
 \begin{subequations}\label{eq.prob1}
 \begin{align}
 \min_{p}& \quad EC_{\rm hom}(p)\\
{\rm s.t.}& \quad \epsilon\leq p \leq C_0.
 \end{align}
 \end{subequations}

As shown in Fig. \ref{fig.curve1}, there is one minimal optimum, denoted by $\hat p$, for $EC_{\rm hom}(p)$, and $\hat p$ may be smaller than $\epsilon$. Formally, we state the following result on traveler's data sharing pricing facing the homogeneous pHs.

 \begin{theorem}\label{Theo1}
 Facing the homogeneous pHs, traveler $\rm T$ should decide the optimal sharing price $p_0^*$ as
 \begin{equation} \label{eq.p_0_star}
 p_0^* = \max\{ \epsilon,\;\hat{p} \},
 \end{equation}
 where $\hat{p}$ is the solution to $\frac{\partial \widetilde{EC}_{\rm hom}(p)}{\partial p}|_{p=\hat{p}}=0$, i.e.,
 \begin{align}\label{eq.hom_derivative}
 & 1-\exp\left(-\lambda\pi d^2\Omega(\hat{p})\right)\Bigg[ 1+ \frac{\lambda\pi d^2(C_0-\hat{p})}{\sqrt{2\pi}\sigma \beta}\notag \\
 & \quad\quad\quad \times \exp\Big(-\Big[\frac{\hat{p}-\epsilon+\beta(Q-B-\mu)}{\sqrt{2}\sigma\beta}\Big]^2\Big)\Bigg]=0.
 \end{align}
 \end{theorem}
 \begin{IEEEproof}
 See Appendix D.
 \end{IEEEproof}

Note that in Theorem~\ref{Theo1}, the optimal solution $p_0^*$ in \eqref{eq.p_0_star} is decided to best tradeoff between the price and the sharing probability. In particular, if $B+\mu\leq Q$, i.e., pH's data quota is larger than the sum of pH's sharing data and the mean data usage, $\hat{p}$ is the unique solution to \eqref{eq.hom_derivative}. Even for this case, it is challenging to obtain $\hat{p}$ in closed-form due to the coupling of variables' coupling in \eqref{eq.hom_derivative}. We apply a bisection search procedure to find the unique $\hat{p}$. On the other hand, if $B+\mu> Q$, the uniqueness of $\hat{p}$ in \eqref{eq.hom_derivative} is not guaranteed. In this case, one can compute the corresponding expected cost under different price values via a one-dimensional exhaustive search, and then select the best solution $\hat{p}$ that returns the minimal expected cost value. Similar to the benchmark case in Section~\ref{sec:Benchmark}, $p_0^*$ decreases with the pH density $\lambda$ and the data quota $Q$ for more data-plan sources, but increases with $\epsilon$, $\mu$, and $\beta$. By checking the relationship between the optimal price $p_0^*$ and pH data usage variance $\sigma^2$, we establish the following proposition.
\begin{proposition}\label{prop.variance}
 If $B+\mu \leq Q$, i.e., the pHs' sharing data amount $B$ plus the expected data usage $\mu$ is within its data quota $Q$, then the traveler's expected cost $EC_{\rm hom}(p)$ increases in the variance $\sigma^2$ of the pHs' data usage.
 \end{proposition}
 \begin{IEEEproof}
 See Appendix E.
 \end{IEEEproof}

Based on Proposition \ref{prop.variance}, as intuitively expected in the case of $B+\mu\leq Q$, an increase of the pH usage variance will increase the chance of data overage beyond quota $Q$. Thus, the traveler needs to pay more to justify such cost increase, and its expected cost increases.

\subsection{Pricing towards pHs of heterogeneous data usage}
In this subsection, we extend the optimal pricing scheme to the case of the heterogeneous pHs. Specifically, we consider a set ${\cal K}\triangleq \{1,\ldots,K\}$ of pH types according to their monthly data usage behaviors (e.g., light- and heavy-usage pH users). Within the same pH type, any two pHs have the same mean usage and variance, and they will subscribe to the same two-part tariff data plan under the WNO.

\subsubsection{The cost analysis for pHs' data sharing}
Now we extend the system model in Section II to the heterogeneous pH case. Denote by $(Q_k,P_{0,k},\beta_k)$, $i\in{\cal K}$, the two-part tariff data plan under the WNO in the $k$th pH type, where $Q_k$ denotes the data quota, $P_{0,k}$ denotes the fixed lump-sum fee, and $\beta_k$ denotes the unit price of the overage data. Suppose again that each pH type in the circular area $A$ of radius $d$ follows an independent PPP with a given spatial density. Specifically, let $\lambda_k$ and ${\cal H}_k$ denote the spatial density and the appearance set of the $k$th pH-type, respectively. Accordingly, $N_k=|{\cal H}_k|$ is a Poisson distributed variable with mean $\lambda_k\pi d^2$. Let $x_{k,i}$ denote the monthly data usage for any pH $i\in{\cal H}_k$ of the $k$th pH-type. Again, we assume that each $x_{k,i}$ follows a Gaussian distribution, i.e., $x_{k,i}\sim{\cal N}(\mu_k,\sigma_k^2)$, where the pH type indices are reordered by accounting for both the mean and variance of pH data-usage, i.e., $0\leq \beta_1(B+\mu_1-Q_1-2\sqrt{2}\sigma_1)\leq \ldots<\beta_K(B+\mu_K-Q_k-2\sqrt{2}\sigma_K)$. Assume that the values $\mu_k$'s and $\sigma_k$'s can be estimated by checking the type-$k$ pHs' previous data usage history.

Define $T_k\triangleq \epsilon+\beta_k(B+\mu_k-Q_k-2\sqrt{2}\sigma_k)$ for any $k\in{\cal K}$. As will be revealed in Theorem 4.2, $T_k$ servers as a threshold price to incentivize any $k$-th pH type, i.e., only when traveler T's offered price $p$ is larger than $T_k$, the pHs of the $k$th type become interested in participating in such data-plan sharing activity. By sharing data of volume $B$, the additional cost of pH $i\in{\cal H}_k$ is expressed as
 \begin{align} \label{eq.cost_overlap}
C_{k,i}(x_{k,i}) =
\begin{cases}
0, &{\rm if}~ 0\leq x_{k,i}\leq Q_k-B \\
\beta_k(x_{k,i}+B-Q_k), &{\rm if}~ Q_k-B\leq x_{k,i}\leq Q_k \\
\beta_k B, &{\rm if}~ x_{k,i}\geq Q_k,
\end{cases}
\end{align}
where $i=1,\ldots,N_k$ and $k=1,\ldots, K$.

\subsubsection{Traveler T's pricing problem}
Under any given ${\cal H}_k$, $k\in{\cal K}$, the probability for traveler T's request to be successfully accepted by pHs is\footnote{Note that $\prod_{i\in{\cal H}_k}{\rm Pr}(p-C_{k,i}<\epsilon) =1$ if ${\cal H}_k=\emptyset$ without any pH presence to possibly share, for $k=1,\ldots,K$.}
\begin{align}\label{eq.het_prob}
 \mathbb{P}_{\tt het}(p|{\cal H}_1,\ldots,{\cal H}_K) &= 1 - \prod_{k\in{\cal K}} \prod_{i\in{\cal H}_k} {\rm Pr}(p-C_{k,i}(x_{k,i})< \epsilon) \notag \\
 &=1-\prod_{k\in{\cal K}}(1-\Omega_k(p))^{N_k},
\end{align}
which takes into account the distribution of any $x_{k,i}$ and $C_{k,i}(x_{k,i})$ in \eqref{eq.cost_overlap}, and we follow a similar analysis as the previous Section~IV.A. Note that in \eqref{eq.het_prob}, the function $\Omega_k(p)$ with respect to $p\in[\epsilon,\beta_kB]$ is defined as
\begin{align}
\Omega_k(p)&\triangleq 1-\frac{1}{2}{\rm erfc}\Bigg(\frac{p-\epsilon+\beta(Q_k-B-\mu_k)}{\sqrt{2}\sigma_k\beta_k}\Bigg) \notag\\
&={\rm Pr}(p-C_{k,i}(x_{k,i})),~~ k\in{\cal K}.
\end{align}
Since that ${\rm erfc}(x)\approx 2$ for $x\leq -2$ and ${\rm erfc}(x)\approx 0$ for $x\geq 2$, it follows that
\begin{align}
\Omega_k(p)
\begin{cases}
=0,& {\rm if}~\frac{p-\epsilon}{\beta_k}+Q_k-B-\mu_k \leq -2\sqrt{2}\sigma_k\\
\in(0,1),&{\rm if}~|\frac{p-\epsilon}{\beta_k}+Q_k-B-\mu_k |\leq 2\sqrt{2}\sigma_k\\
=1,&{\rm if}~\frac{p-\epsilon}{\beta_k}+Q_k-B -\mu_k \geq 2\sqrt{2}\sigma_k,
\end{cases}
\end{align}
for any $k\in{\cal K}$, where $\epsilon\leq p\leq \epsilon+\beta B$.

 \begin{figure}
 \centering
 \includegraphics[width = 3.0in]{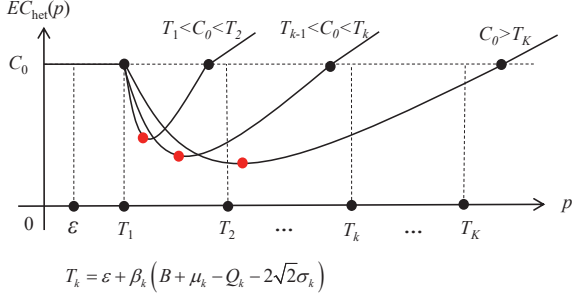}
 \caption{An illustration of function $EC_{\rm het}(p)$ for $K$ types of heterogeneous pHs.} \label{fig.Het}
 \end{figure}

Considering all the possible pH sets ${\cal H}_k$'s, the probability for traveler T's data sharing request to be successfully accepted by a pH is then
\begin{align}\label{eq.P_het}
 & \mathbb{P}_{\tt het}(p)= 1-\prod_{k=1}^K \sum_{N_k=0}^\infty (1-\Omega_k(p))^{N_k}\frac{(\lambda_k\pi d^2)^{N_k}}{N_k!}e^{-\lambda_k\pi d^2}
\notag \\
&= \begin{cases}
0,&{\rm if}~ p\leq T_1\\
1-\exp(-\sum_{j=1}^k \lambda_j\pi d^2\Omega_j(p)),&{\rm if}~ T_k\leq p\leq T_{k+1},\\
&\quad~~k=1,\ldots,K-1\\
1-\exp(-\sum_{j=1}^K \lambda_j\pi d^2\Omega_j(p)),&{\rm if}~p\geq T_K.
\end{cases}
\end{align}
As a result, traveler T's expected cost $EC_{\rm het}(p)$ in data sharing is expressed as
\begin{align}\label{eq.EC2}
 & EC_{\rm het}(p)  = p\times \mathbb{P}_{\tt het}(p) + C_0\times \big(1-\mathbb{P}_{\tt het}(p) \big)= \notag\\
 & \begin{cases}
C_0,&{\rm if}~ p\leq T_1\\
C_0+(p-C_0)(1-e^{-\sum_{j=1}^k \lambda_j\pi d^2\Omega_j(p)}),&{\rm if}~ T_k\leq p\leq T_{k+1},
\\
& k=1,\ldots,K-1\\
C_0+(p-C_0)(1-e^{-\sum_{j=1}^K \lambda_j\pi d^2\Omega_j(p)}),&{\rm if}~p\geq T_K.
\end{cases}
 \end{align}
 We next formulate the traveler T's expected cost minimization problem as
 \begin{subequations}\label{eq.prob2}
 \begin{align}
 \min_{p}& ~~ EC_{\rm het}(p)\\
{\rm s.t.}& ~~ \epsilon\leq p \leq C_0,
 \end{align}
 \end{subequations}
which is more involved than problem \eqref{eq.prob1} with homogeneous pHs. Note that $EC_{\rm het}(p)$ is a piecewise convex function of $p$ under the condition of $B+\mu_k\leq Q_k$ for any $k\in{\cal K}$ (see appendix F). Denote $p^*_{\rm het}$ as the optimal solution to \eqref{eq.prob2}.

To solve \eqref{eq.prob2}, we first denote $p^*_k$ as the solution of $\frac{\partial g_k(p)}{\partial p}|_{p=p_k^*}=0$, $k\in{\cal K}$, i.e.,
 \begin{align}\label{eq.p_k}
& 1-\exp\Big(-\sum_{j=1}^k\lambda_j\pi d^2\Omega_j(p)\Big)\Bigg[1+\sum_{j=1}^k\frac{\lambda_j\pi d^2(C_0-p)}{\sqrt{2\pi}\sigma_j\beta_j}\notag \\
&\quad \times \exp\Big(-(\frac{p-\epsilon+\beta_j(Q_j-B-\mu_j)}{\sqrt{2}\sigma_j\beta_j})^2\Big)\Bigg]=0,
\end{align}
where $g_k(p)\triangleq C_0+(p-C_0)\Big(1-\exp\big(-\sum_{j=1}^k \lambda_j\pi d^2\Omega_j(p)\big)\Big)$, $k\in{\cal K}$. Note that solving \eqref{eq.p_k} for the heterogeneous pH case is also involved as solving \eqref{eq.hom_derivative} for the homogeneous pH case, and it is challenging to obtain the optimal $p_k^*$ in a closed form. Similarly, one can turn to numerical approach for searching for the optimal solution $p_k^*$. Fig.~\ref{fig.Het} illustrates the possible cost curves of the traveler T's expected cost $EC_{\rm het}(p)$ for the $K$ types of heterogenous pHs under different values of $C_0$.

Note that the first-type pHs (i.e., light pHs) are more willing to accept the data-plan sharing request from the traveler than the second type pHs (i.e., heavy pHs) since that the light pHs admit smaller data-sharing cost than the heavy ones. When $C_0$ is large, traveler T tries to attract the first $k$ types of pHs in data-plan sharing, by designing a large price $p_{k}^*$. It is shown that in the case of $C_0\leq T_k$, $k\in{\cal K}$, the heavy pHs from the $k$th to the $K$th types will not consider data sharing due to higher cost than roaming fee.

Formally, we now establish the theorem on the optimal sharing price $p_{\tt het}$ for traveler T to minimize his expected cost to attract heterogeneous pHs as follows.
\begin{theorem}\label{Theo2}
Facing the heterogeneous $K$ pH types, traveler T decides the optimal sharing price $p^*_{\tt het}$ by targeting different groups of pHs:
 \begin{itemize}
 \item For $C_0\leq T_1$, i.e., $C_0\leq \epsilon+\beta_1\big(B+\mu_1-Q_1-2\sqrt{2}\sigma_1\big)$, then the optimal price is $p^*_{\tt het}=C_0$, since none of pHs will accept the data-plan sharing and traveler T has to resort to the WNO.
\item For $T_k\leq C_0 \leq T_{k+1}$, i.e., $\epsilon+\beta_k\big(B+\mu_k-Q_k-2\sqrt{2}\sigma_k\big)\leq C_0 \leq \epsilon+\beta_{k+1}\big(B+\mu_{k+1}-Q_{k+1}-2\sqrt{2}\sigma_{k+1}\big)$, for any $k=1,\ldots,K-1$, then the optimal price to minimize $EC_{\tt het}(p)$ is $p^*_{\tt het}=p^*_k$. This implies that traveler T targets for the first $k$ types of pHs in data sharing.
\item For $C_0\geq T_K$, i.e., $C_0\geq \epsilon+\beta_K\big(B+\mu_K-Q_K-2\sqrt{2}\sigma_K\big)$, then all the $K$ pH types are involved in data-plan sharing and the optimal price is $p^*_{\tt het}=p^*_K$.
\end{itemize}
 \end{theorem}
 \begin{IEEEproof}
 See Appendix F.
 \end{IEEEproof}

Theorem~\ref{Theo2} tells a threshold sharing price $T_k$ to incentivize the pHs of the $k$th type. When the roaming fee $C_0$ for the traveler is smaller than the threshold, the traveler should design such a reward to attract (or incentivize) the pHs of the first $(k-1)$th pH types. On the other hand, when the roaming fee $C_0$ is still larger than the threshold, the traveller will refer to $T_{k+1}$ for the optimal price targeting at the first $k$ pH types.

\section{Extension of Pricing for Overlapped Travelers}\label{sec:Multiple}
In this section, we consider the coexistence of possibly multiple travelers in the area $A$ of radius $d$, where multiple travelers request data-plan sharing from the common pool of pHs. On the behalf of all travelers in this pH-sharing platform, we aim to study the optimal pricing for all travelers to minimize their total expected costs in pH data-plan sharing. Intuitively, as there are more travelers, we need to decide a larger uniform price in the data-plan sharing platform for attracting more pHs to share.

We focus on a typical traveler $\rm T$ in the circle area $A$ of radius $d$. He also face a random number of other $M\geq 0$ travelers requesting pH connection. To capture the mobility of these travelers, we assume that the number $M$ of overlapped travelers follows an independent PPP with spatial density $\lambda_t>0$. The PMF of $M$ is then given by
 \begin{align}\label{eq.M}
 {\rm Pr}(M=m) = \frac{(\lambda_t\pi d^2)^m}{m!}\exp\left(-\lambda_t\pi d^2\right),~~m=0,1,\cdots
 \end{align}
The data sharing scheme in Fig.~\ref{fig.protocol} also applies here, where each traveler offers a uniform price $p$ decided centrally by the sharing platform to pHs and we consider that each pH $i\in{\cal H}$ with $p-C_i(x_i)\geq \epsilon$ can reply to serve at most one traveler due to capacity limit.

\subsection{Analysis of pHs' sharing probability}
For ease of exposition, we consider homogeneous pH case and an identical two-part tariff data plan for all the pHs in pH set $\cal H$ in this section. Denote by $N_y$ the random number of pHs satisfying the participation condition of $p-C_i(x_i)\geq \epsilon$, where $N_y\leq N$. Conditioned on the presence of arbitrary $N_y$ pHs in the vicinity, the probability for traveler T to be accepted and served by pHs is ${\rm min}\{1,N_y/(M+1)\}$. The ``min'' operation is adopted to ensure that the accepted probability is equal to 1 in the case of $N_y>M+1$. As a result, given $N$ and $M$ in the area $A$ of our interest, the probability for the overlapping traveler T to be served successfully is
\begin{align} \label{eq.mul_Pr}
 &\mathbb{P}_{\tt mul}(p | N,M)=\notag \\
 &\begin{cases}
0,\quad\quad\quad\quad\quad\quad\quad\quad\quad\quad\quad\quad\quad\quad\quad\quad\quad\quad  N=0\\
\sum_{N_y=1}^{N} \min\left\{1,\frac{N_y}{M+1}\right\}{N\choose N_y}\left(\Omega(p)\right)^{N_y}\left(1-\Omega(p)\right)^{N-N_y},\\ \quad\quad \quad\quad\quad\quad\quad\quad\quad\quad\quad\quad\quad\quad\quad\quad\quad\quad\quad N\geq 1,
\end{cases}
\end{align}
where $\Omega(p)={\rm Pr}(p-C_i(x_i)\geq \epsilon)$ as in \eqref{eq.prob_A}. In the following, we focus on the reasonable range $[\epsilon,\beta B]$ of optimal $p$, and omit the two trivial cases of ${\rm Pr}(p-C_i(x_i)\geq \epsilon)=0$ and ${\rm Pr}(p-C_i(x_i)\geq \epsilon)=1$ for $p<\epsilon$ and $p>\beta B$, respectively.

Taking expectation of \eqref{eq.mul_Pr} over any possible pH number $N$, we denote by $\mathbb{P}_{\tt mul}(p|M)$ the probability for traveler T to establish the pH connection in the area $A$ under the given $M$ travelers nearby. Readily, it follows that
\begin{align}\label{eq.A_cond}
\mathbb{P}_{\tt mul}(p|M)&=\sum_{N=0}^{\infty}\mathbb{P}_{\tt mul}(p|N,M)\frac{(\lambda\pi d^2)^N \exp\left(-\lambda\pi d^2\right)}{N!}\notag \\
&=\sum_{N=1}^{\infty}\sum_{N_y=1}^{N} \min\left\{1,~\frac{N_y}{M+1}\right\}{N\choose N_y} (\Omega(p))^{N_y}\notag \\
&\quad \times (1-\Omega(p))^{N-N_y} \frac{(\lambda\pi d^2)^{N} \exp\left(-\lambda\pi d^2\right)}{N!}.
\end{align}
Taking further expectation of $\mathbb{P}_{\rm mul}(p|M)$ over the random overlapped traveler number $M$, the successful pH data-sharing probability for traveler T is given by
\begin{align}\label{eq.f}
\mathbb{P}_{\tt mul}(p) =\sum_{M=0}^{\infty}\mathbb{P}_{\tt mul}(p|M)\frac{(\lambda_t\pi d^2)^M\exp\left(-\lambda_t\pi d^2\right)}{M!},
\end{align}
which is difficult to simplify due to the operation $\min \{1,N_y/(M+1)\}$ in each summation term.

 \begin{figure}
 \begin{center}
 \includegraphics[width=2.8in]{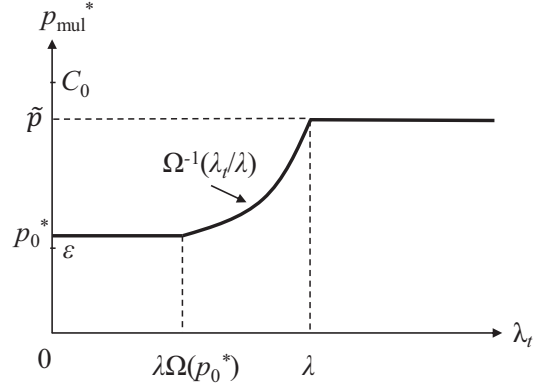}
 \end{center}
 \caption{An illustration of the optimal price $p_{\rm mul}^*$ with respect to the nearby travelers' density $\lambda_t$ in approximate problem \eqref{eq.prob-multi-traveler}.}\label{fig.p_mul}
 \end{figure}

 Readily, it holds that $\min\{1,N_y/(M+1)\} \leq 1$ and $\min\{1,N_y/M+1\}\leq N_y/(M+1)$ for any possible $N_y$ value in the summation \eqref{eq.A_cond}. Then, the original cost objective in \eqref{eq.f} now increases to two upper bounds $\mathbb{P}^{\tt UB1}_{\tt mul}(p)$ and $\mathbb{P}^{\tt UB2}_{\tt mul}(p)$ via these two replacements for any $N_y$ value, respectively. Otherwise, the discontinuous ``min'' operation kept for any possible $N_y$ value makes \eqref{eq.f} difficult to analyze and optimize. Specifically, the first upper bound $\mathbb{P}^{\tt UB1}_{\tt mul}(p)$ is obtained by choosing $\min\{1, N_y/(M+1)\}=1$ in each summation term in \eqref{eq.mul_Pr},
i.e.,
 \begin{align}\label{eq.P_UB1}
  \mathbb{P}^{\tt UB1}_{\tt mul}(p)  &\triangleq
 \sum_{M=0}^{\infty}\sum_{N=1}^{\infty}\sum_{N_y=1}^{N}{N\choose N_y}(\Omega(p))^{N_y}(1-\Omega(p))^{N-N_y} \notag \\
 &\quad \times \frac{(\lambda\pi d^2)^{N} \exp(-\lambda\pi d^2)}{N!}\frac{(\lambda_t\pi d^2)^M\exp(-\lambda_t\pi d^2)}{M!}\notag \\
  &=1-\exp\big(- \lambda\Omega(p)\pi d^2\big).
 \end{align}
In \eqref{eq.P_UB1}, it is assumed that a low traveler density case (i.e., $\lambda_t$ is small) where pH supply is sufficient for traveler demand. On the other hand, for the high traveler density case, the second upper bound $\mathbb{P}_{\tt mul}^{\tt UB2}(p)$ is obtained, by choosing $\min\{1,N_y/(M+1)\}=N_y/(M+1)$ in each summation term in \eqref{eq.mul_Pr}, i.e.,
 \begin{align}\label{eq.P_UB2}
 \mathbb{P}^{\tt UB2}_{\tt mul}(p) & \triangleq  \sum_{M=0}^{\infty}\sum_{N=1}^{\infty}\frac{\sum_{N_y=1}^{N} N_y{N\choose N_y}(\Omega(p))^{N_y}(1-\Omega(p))^{N-N_y}}{M+1}\notag \\
 &\quad \times \frac{(\lambda\pi d^2)^N \exp(-\lambda\pi d^2)}{N!}\frac{(\lambda_t\pi d^2)^M\exp(-\lambda_t\pi d^2)}{M!} \notag \\
 &= \frac{\lambda \Omega(p)\left(1-\exp(-\lambda_t\pi d^2)\right)}{\lambda_t}.
 \end{align}
In \eqref{eq.P_UB2}, it is assumed that pH supply is not enough to meet the traveler demand. Note that the detailed derivations of these two upper bounds are relegated to Appendix G. We denote $\lambda_t^*$ as the solution of $\mathbb{P}^{\tt UB1}_{\tt mul}(p)= \mathbb{P}^{\tt UB2}_{\tt mul}(p)$, i.e.,
\begin{align}\label{eq.lambda_t_star}
1-\exp(- \lambda\Omega(p)\pi d^2) = \frac{\lambda \Omega(p)\left(1-\exp(-\lambda^*_t\pi d^2)\right)}{\lambda^*_t}.
\end{align}
It is verified that $\lambda_t^*=\lambda\Omega(p)$ in \eqref{eq.lambda_t_star}, which implies that the two upper bounds $\mathbb{P}^{\tt UB1}_{\tt mul}(p)$ and $\mathbb{P}^{\tt UB2}_{\tt mul}(p)$ become identical when the traveler density $\lambda_t$ is equal to the expected density $\lambda\Omega(p)$ of pHs who are willing to share data.
Since that the pH sharing probability $\Omega(p)$ in \eqref{eq.prob_A} increases with $p$, we jointly propose a tighter upper bound of $\mathbb{P}_{\tt mul}^{\tt UB}(p)=\min\{\mathbb{P}^{\tt UB1}_{\tt mul}(p), \mathbb{P}^{\tt UB2}_{\tt mul}(p)\}$. In particular, we approximate $\mathbb{P}_{\tt mul}(p)$ in \eqref{eq.f} as
 \begin{align}\label{eq.P_mul_UB}
 \mathbb{P}_{\tt mul}^{\tt UB}(p)=\begin{cases}
 1-\exp( -\lambda \Omega(p)\pi d^2), &{\rm if}~\lambda_t < \lambda\Omega(p),\\
 \frac{\lambda \Omega(p)(1-\exp(-\lambda_t\pi d^2))}{\lambda_t} &{\rm if}~\lambda_t\geq \lambda\Omega(p).
 \end{cases}
 \end{align}

\subsection{Traveler T's pricing for cost minimization}
We aim to minimize the typical traveler T's expected cost, i.e.,
$EC(p)=\mathbb{P}_{\tt mul}(p)\times p+ (1-\mathbb{P}_{\tt mul}(p))\times C_0$, which is equivalent to minimizing all the $(M+1)$ travelers' total expected cost. As $\mathbb{P}_{\tt mul}(p)$ and $EC(p)$ are not tractable, we turn to an approximate optimization problem by using $\mathbb{P}_{\tt mul}^{\tt UB}(p)$ in \eqref{eq.P_mul_UB} and a lower bound of $EC(p)$:
\begin{subequations}\label{eq.prob-multi-traveler}
\begin{align}
\min_{p} &~ \mathbb{P}_{\tt mul}^{\tt UB}(p)\times p + (1-\mathbb{P}_{\tt mul}^{\tt UB}(p))\times C_0\\
{\rm s.t.}&~ \epsilon\leq p \leq C_0.
\end{align}
\end{subequations}
Denote by $p^*_{\tt mul}$ the optimal solution to the approximate problem \eqref{eq.prob-multi-traveler}. Since $\mathbb{P}_{\tt mul}^{\tt UB}(p)$ depends on the nearby traveler density $\lambda_t$, we next discuss the following three regimes for solving \eqref{eq.prob-multi-traveler}.
\subsubsection{Low traveler density regime ($0\leq \lambda_t\leq \lambda \Omega(p_0^*)$ with $p_0^*$ given in Theorem \ref{Theo1})}
This regime includes sufficient pH supply for travelers, and is similar to Theorem \ref{Theo1} without traveler overlap. The traveler T only needs to consider the pH response without coordination with other travelers. In this case, $\mathbb{P}_{\tt mul}^{\tt UB}(p)=\mathbb{P}_{\tt mul}^{\tt UB1}(p)$ and $p_0^*$ is the optimal solution to problem \eqref{eq.prob-multi-traveler}. Note that $p_{\rm mul}^*=p_0^*$ is independent of $\lambda_t$ as shown in Fig.~\ref{fig.p_mul}.

\subsubsection{Medium traveler density regime ($\lambda \Omega(p_0^*)\leq \lambda_t \leq \lambda)$}
In this regime, the average traveler demand is more than the shared pH supply at traditional price $p_0^*$, and the traveler T needs to increase price $p$ for minimizing $\mathbb{P}_{\tt mul}^{\tt UB2}(p)\times p + (1-\mathbb{P}_{\tt mul}^{\tt UB2}(p))\times C_0$ until $\lambda_t=\lambda\Omega(p)$.
\begin{lemma}\label{lem.mediate}
Facing a medium traveler density regime (i.e., $\lambda \Omega(p_0^*)\leq \lambda_t \leq \lambda$), the optimal data sharing price $p_{\rm mul}^*$ for \eqref{eq.prob-multi-traveler} is
\begin{align}
p^*_{\rm mul}=\Omega^{-1}(\lambda_t/\lambda),
\end{align}
where $\Omega^{-1}(x)$ is an inverse function of $\Omega(x)$ in \eqref{eq.prob_A}. This price $p_{\rm mul}^*$ decreases in the pH density $\lambda$ and increases in the overlapped traveler density $\lambda_t$ (see Fig. \ref{fig.p_mul}).
\end{lemma}
\begin{IEEEproof}
See Appendix H.
\end{IEEEproof}

\subsubsection{High traveler density regime ($\lambda_t>\lambda$)}
In this regime, it always holds that $\lambda_t> \lambda \Omega(p)$ due to $\Omega(p)<1$ and the pH supply is sufficient to meet travelers' demand. The approximated cost for traveler T is always $\mathbb{P}_{\tt mul}^{\tt UB2}(p)\times p + (1-\mathbb{P}_{\tt mul}^{\tt UB2}(p))\times C_0$, which is given by
\begin{align}
EC_{\rm high}(p)\triangleq C_0+(p-C_0) \frac{\lambda \Omega(p)(1-\exp(-\lambda_t\pi d^2))}{\lambda_t}.
\end{align}

\begin{lemma}\label{lem.high}
Facing a high traveler density regime (i.e., $\lambda_t>\lambda$), the optimal price to minimize the expected cost is obtained as $p_{\tt mul}^*=\tilde{p}$ (see Fig. \ref{fig.p_mul}), where $\tilde{p}$ is independent of $\lambda_t$ and is the unique solution to $\frac{\partial EC_{\rm high}(p)}{\partial p}|_{p=\tilde{p}}=0$, i.e.,
\begin{align} \label{eq.p_tilde}
&1-\frac{1}{2}{\rm erfc}\Big(\frac{\tilde{p}-\epsilon+\beta(Q-B-\mu)}{\sqrt{2}\sigma\beta}\Big)
 \notag\\
 &\quad \quad +\frac{\tilde{p}-C_0}{\sqrt{2\pi}\sigma\beta}\exp\Big(-\Big[\frac{\tilde{p}-\epsilon+\beta(Q-B-\mu)}{\sqrt{2}\sigma\beta}\Big]^2\Big)=0.
\end{align}
\end{lemma}

 Note that it is challenging to obtain the closed-form of the solution $\tilde{p}$ to \eqref{eq.p_tilde}. Again, one resorts to numerically find the value of $\tilde{p}$ by implementing a bisectional search procedure.

By summarizing the above results, we have the following proposition.

\begin{proposition}\label{theo.mul}
In the overlapped traveler pricing problem in \eqref{eq.prob-multi-traveler}, the optimal price $p_{\rm mul}^*$ is
\begin{align}
p_{\rm mul}^* = \begin{cases}
p_0^* ~{\rm{ in}}~ \eqref{eq.p_0_star},&~{\rm if}~0\leq \lambda_t\leq \lambda \Omega(p_0^*)\\
\Omega^{-1}(\frac{\lambda_t}{\lambda}),&~{\rm if}~\lambda \Omega(p_0^*) < \lambda_t \leq \lambda \\
\tilde{p} ~\rm{in}~ \eqref{eq.p_tilde}, &~{\rm if}~\lambda_t>\lambda.
\end{cases}
\end{align}
\end{proposition}

\subsection{Extension to the case for one pH sharing data with multiple travelers}
Suppose that the support number of a pH for sharing is $q\geq 1$. Then we need to consider all the possible matchings between all pH connections and travelers in the same area and there are many possible connection-traveler combinations, as not $q$ connections of a pH will be shared successfully. For a pH, its final estimation of sharing cost needs to aggregate all the possible costs incurred by all the connections subject to the same monthly data quota, and such quota-sharing correlation among connections is very involved for computing the total cost. Alternatively, we propose a more tractable way via decomposition approximation of such connections, by treating each pH $i\in{\cal H}$ as a number $q$ of identical and independent sub-pHs. Each sub-pH $j_i\in \{i(1),\ldots,i(q)\}$ is then subscribed to an equally partitioned data-plan of $(Q_i/q, P_0/q, \beta_i)$ and only cares for its own cost for sharing decision. Then for each sub-pH $j_i$, the data sharing cost and sharing probability can be obtained similarly as \eqref{eq.cost_overlap} and \eqref{eq.A_cond}, respectively. Then we can still derive the optimal pricing by just considering the $q$ sub-pHs' independent responses as in Theorem 4.2 for final pricing computation. It is interesting to note that a pH $i$ (especially facing a big data request of amount $B$ from the traveler) may not want to activate all $q$ connections due to thinned quota $Q_i/q$ of each connection to meet $B$. That is, pH $i\in{\cal H}$ may only open a subset of connections to center the data resource.

 \section{Numerical Results}\label{sec:Numerical}

 \subsection{Monopoly Traveler Case}
 In this subsection, we evaluate the proposed incentive pricing scheme in the homogeneous and the heterogeneous pH cases, as well as the benchmark cost for the social optimum under complete information. In the simulations, the pH range is set as $d=30$~meters (m).

 \begin{figure}
 \centering
 \includegraphics[width = 3.4in]{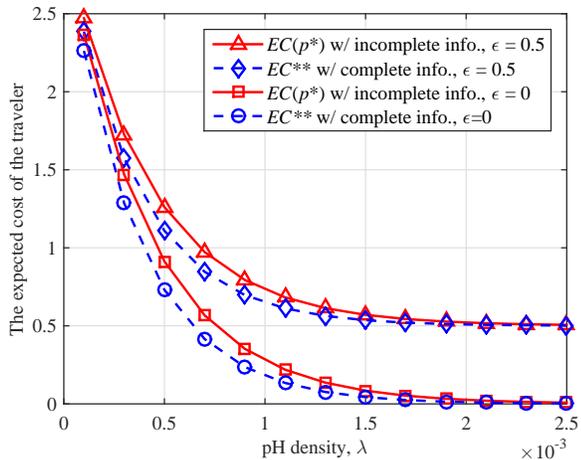}\\
 \caption{The traveler's expected cost versus the homogeneous pH density $\lambda$ under different reservation utility $\epsilon$ values.} \label{fig.vs_zeta}
\end{figure}

Fig.~\ref{fig.vs_zeta} shows the traveler T's expected cost versus the homogeneous pH spatial density $\lambda$ under different reservation utility values, where $C_0=\$3$, $B=0.2$~GB, $\mu=1.7$~GB, and $\sigma^2=0.01$. The same two-part tariff plan $(2~\text{GB},\$17,\$0.013/\text{MB})$ is set for all the pHs. It is corroborated that the social optimum cost $EC^{**}$ under complete information is a lower bound for the proposed ones under incomplete information, and the cost gap becomes smaller as $\lambda$ increases. It is observed in Fig.~\ref{fig.vs_zeta} that the traveler T's expected cost decreases as $\lambda$ increases, close to the reservation utility $\epsilon$ under either complete or incomplete information. As expected, the sharing price is below bounded by $\epsilon$, and a larger $\lambda$ implies more pHs nearby, which leads to a high probability for the traveler to incentivize pHs at a lower sharing price. 

\begin{figure}
\begin{center}
\includegraphics[width=3.4in]{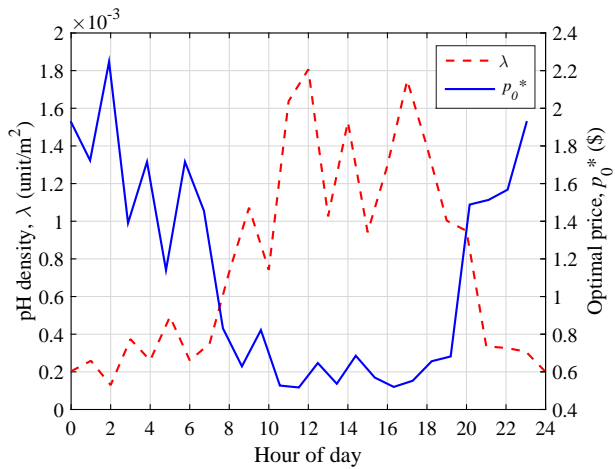}
\end{center}
\caption{The optimal price announced by traveler T versus the pH density on an hour-of-day basis.} \label{fig.lambda_dynamic}
\end{figure}

Fig.~\ref{fig.lambda_dynamic} shows the optimal price $p_0^*$ announced by traveler T versus the pH density $\lambda$ on an hour-of-day basis. By following the usuage patterns in \cite{Will09}, during the night (9 p.m.--7 a.m.), we set the pH density $\lambda$ to follow a uniform distribution $\lambda\in[0.1,0.5]\times 10^{-3}$ unit per square meter (unit/m$^2$), and for the day (8 a.m.--8 p.m.), the pH density is set to follow a uniform distribution $\lambda\in[0.5,2]\times 10^{-3}$ unit/m$^2$. The pHs' reservation utility is set as $\epsilon=0.5$ and the other parameters are identical to those in Fig.~\ref{fig.vs_zeta}. It is observed in Fig.~\ref{fig.lambda_dynamic} that a larger pH density $\lambda$ value corresponds to a smaller price $p_0^*$ value, and vice versa. This is expected since that the traveler could benefit from the data resource from a large number of pHs in the sharing area, thereby offering a smaller price for data sharing. In addition, Fig.~\ref{fig.lambda_dynamic} shows two distinct periods roughly corresponding to day and night, and have low and high prices for traveler T. This is expected since that a significantly high pH density appears in day time than that in night for sharing with the traveler.

 \begin{figure}
 \centering
 \includegraphics[width = 3.4in]{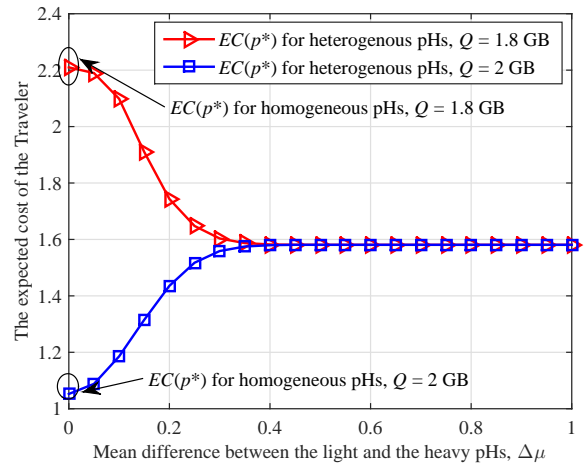}\\
\caption{The traveler T's expected cost versus $\Delta\mu$ under different monthly data quotas for pHs.} \label{fig.vs_delta_mu}
\end{figure}

 Fig.~\ref{fig.vs_delta_mu} shows the traveler T's expected cost versus the mean difference $\Delta\mu$ between the light and the heavy pHs under different monthly data quotas for all the pHs, where $\mu_1 = \mu - \Delta\mu/2$, $\mu_2 = \mu+\Delta\mu/2$, the light and the heavy pH densities are $\lambda_1=\lambda_2=\lambda/2=2.5\times10^{-4}$, the reservation utility is $\epsilon=\$0.2$, and the remaining parameters are set the same as those in Fig.~\ref{fig.vs_zeta}. It is observed that the expected cost is about $EC_{\rm het}(p^*)=\$2.2$ for the homogeneous pHs at $Q=1.8$~GB, while the expected cost $EC_{\rm het}(p^*)$ for the heterogeneous pHs first decreases in $\Delta\mu$ and then remains unchanged. By contrast, at $Q=2$~GB, the expected cost $EC_{\rm hom}(p^*)$ is about $\$1.05$ for the homogeneous pHs, while $EC_{\rm het}(p^*)$ first increases in $\Delta\mu$ and then remains unchanged for the heterogenous pHs. This is expected that, when pHs with a limited data quota (e.g., small $Q$), the traveler could benefit from the diverse of the pH data usage since the sharing cost reduces for the light pHs; on the other hand, with a large data quota, the traveler may not benefit from the diverse of the pH data usage due to the increasing of the sharing cost for the heavy pHs.

It is interesting to observe in Fig.~\ref{fig.vs_delta_mu} that, when $\Delta \mu$ becomes large, both the traveler's expected cost with $Q=1.8$~GB and that with $Q=2$~GB converge to a fixed value $EC_{\rm het}(p^*)=\$1.58$ for the heterogeneous pHs. The reason is as follows. For the very diverse heterogeneous pHs (i.e., when $\Delta\mu$ becomes large), the traveler rewards the light pHs with a fixed price $\epsilon$ and the heavy pHs with a fixed price $(\epsilon+\beta B)$, respectively; thus the traveler's expected cost is independent with the pH's monthly data quota $Q$ in this case.

\subsection{Overlapping Travelers Case}
In this subsection, we consider pH data sharing in the overlapping travelers case. Specifically, the identical two-part tariff plan $(2{\rm GB}, \$17,\$0.013/{\rm MB})$ and the reserved utility $\epsilon=\$0.5$ are set for all the pHs in the area $A$. The data usage statistics are also set to be identical for all the pHs with mean $\mu=1.8$ GB and variance $\sigma^2=0.01$. The data roaming fee is set to be $C_0=\$3$ and the data usage amount $B=0.29$ GB for each traveler. We set the pH-range and the pHs' spatial density as $d=30$ m and $\lambda=10^{-3}$ unit/m$^2$, respectively.

\begin{figure}
\begin{center}
\includegraphics[width=3.4in]{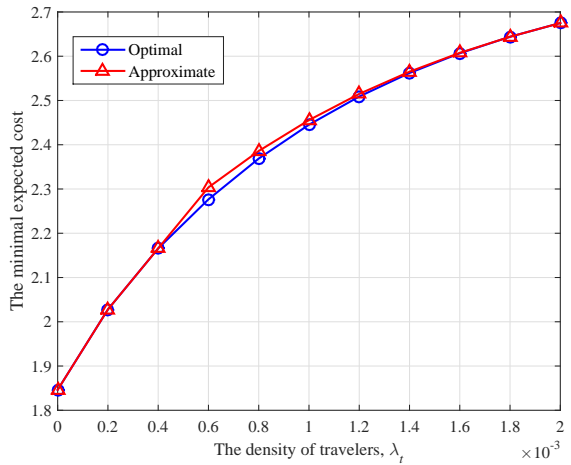}
\end{center}
\caption{The minimal expected cost $EC(p)$ for traveler $\rm T$ in pH data sharing versus the nearby travelers' density $\lambda_t$.}\label{fig.multiuser_cost_vs_lambda_t}
\end{figure}

Fig.~\ref{fig.multiuser_cost_vs_lambda_t} shows the expected cost for traveler T in pH data sharing versus the nearby travelers' density $\lambda_t$. It is observed that both the optimal and the approximate expected costs increase as $\lambda_t$ increases. It is also observed that in Fig.~\ref{fig.multiuser_cost_vs_lambda_t}, when $4\times10^{-4} \leq \lambda_t \leq 10^{-3}$, the approximate expected cost based on the $\mathbb{P}_{\tt mul}^{\tt UB}(p)$ is slightly higher than the optimal one; in the small and large $\lambda_t$ regimes, the approximate approach achieves a close performance as the optimal one. It suggests that one can safely evaluate the minimal expected cost with $\mathbb{P}^{\tt UB}_{\tt mul}(p)$ in the overlapping travelers case with a high traveler density.

\section{Conclusions}\label{sec:Conclusion}

In this paper, we investigated the optimal reward-based pricing problem in opportunistic pH data-plan sharing to reach a win-win situation. It was established that the benchmark social optimum under complete information. Taking into account the random mobility and the cost distribution of selfish pHs nearby, the optimal price was derived to minimize the traveler's expected cost facing the homogeneous pHs, and further extended to the case of heterogeneous pHs. Lacking selfish pHs' information and cooperation, the traveler's expected cost is higher than that under the complete information, but the gap diminishes as the pH spatial density increases. It was shown that the traveler may or may not benefit from the diversity of pHs' data usage behavior. As an extension, the overlapped-traveler pricing is investigated by resorting to a tractable lower bound approximation. The (near-)optimal sharing price is efficiently determined and serves as a safe approximation especially in large pH density case. We show that the optimal price increases with the density of travelers. Numerical results evaluated the effectiveness of the proposed schemes.

As a future direction, we will pursue a more general pH data sharing pricing scheme by allowing the traveler to adjust his required data amount $B$ at the same time for the elastic traffic applications (such as web browsing)\cite{mus2006wifi}. Another interesting research direction for pH data-plan sharing in roaming market is based on auction mechanisms (e.g., Vickrey-Clarke-Groves (VCG) auction), where pHs (if any) first announce their costs to the traveler and then traveler decides to activate which pH for data sharing. 

\section*{Appendix}
\subsection{Proof of Lemma \ref{lem.CDF}}\label{lem.CDF_proof}

First consider the case of $c=\beta B$. According to \eqref{eq.cdf_Ci}, the range of $C_i(x_i)$ is $[0,\beta B]$. Hence, it always holds that $C_i(x_i)\leq \beta B$, i.e., ${\rm Pr}(C_i(x_i)\leq \beta B) = 1$.

Next, based on \eqref{eq.cdf_Ci}, if $0\leq x_i\leq Q-B$, then $C_i(x_i)=0$ and we have
\begin{align} \label{eq.c0}
{\rm Pr}(C_i(x_i)=0)&={\rm Pr}(0\leq x_i\leq Q-B)\notag \\
&= \int_{0}^{Q-B}\Phi(x_i)dx_i \notag \\
&=\Phi(Q-B).
\end{align}
According to \eqref{eq.cdf_Ci}, for $Q-B<x_i<Q$, it holds that $C_i(x_i)=\beta(x_i+B-Q)$. By setting $C_i(x_i)=c$ for $0<c<\beta B$, we have $x_i=c/\beta+Q-B$. Hence, under given $0<c<\beta B$, it follows that
\begin{align}\label{eq.cn0}
{\rm Pr}(C_i(x_i)\leq c) &= {\rm Pr}\big(x_i<\frac{c}{\beta}+Q-B\big)\notag \\
&=\int_{0}^{\frac{c}{\beta}+Q-B}\Phi(x_i)dx_i \notag \\
&= \Phi\big(\frac{c}{\beta}+Q-B\big).
\end{align}
Based on \eqref{eq.c0} and \eqref{eq.cn0}, the probability of $C_i(x_i)=0$ becomes identical to that of $C_i(x_i)\leq c$ for $0<c<\beta B$. Hence, it is immediate that ${\rm Pr}(C_i(x_i)\leq c) = \Phi(\frac{c}{\beta}+Q-B)$ for $0\leq c <\beta B$. Until now, Lemma~\ref{lem.CDF} is proved.

\subsection{Proof of Lemma \ref{lem.c1}}
Under the given pH set, it is required to obtain the expected value of $p(N)=\epsilon+\mathbb{E}\{C_1\}$, where the expectation is taken over any possible $C_1$. With \eqref{eq.cdf_c1} and \eqref{eq.pdf_c1}, we derive $\mathbb{E}\{C_1\}$:
\begin{align} \label{eq.EC1}
&\mathbb{E}\{C_1\} = \int_{0}^{\beta B}xf_{C_1}(x)dx+\beta B\times {\rm Pr}(C_1=\beta B)\notag \\
&=\int_{0}^{\beta B}x d\left[1-(1-\Phi(\frac{x}{\beta}+Q-B))^N\right] + \beta B(1-\Phi(Q))^N\notag \\
&=x[1-(1-\Phi(\frac{x}{\beta}+Q-B))^N]\Big|_{0}^{\beta B} \notag \\
&\quad - \int_{0}^{\beta B}[1-(1-\Phi(\frac{x}{\beta}+Q-B))^N]dx+\beta B(1-\Phi(Q))^N \notag \\
&= \beta B\left[1-(1-\Phi(Q))^N\right] -\beta B \notag \\
&\quad +\int_{0}^{\beta B}(1-\Phi(\frac{x}{\beta}+Q-B))^N dx + \beta B(1-\Phi(Q))^N \notag \\
&=\int_{0}^{\beta B}(1-\Phi(\frac{x}{\beta}+Q-B))^N dx,
\end{align}
where the third equality of \eqref{eq.EC1} holds by following the integration by parts. It thus follows Lemma \ref{lem.c1} and the expected cost of traveler T under given $N\geq 1$ pHs is given by \eqref{eq.lem1}.

\subsection{Proof of Theorem \ref{theo.c1}}
Based on Lemma \ref{lem.c1} and considering all the possibilities of $N$ (including $N=0$), the expected cost for traveler T is given by\cite{Fishburn70}
\begin{align}\label{eq.EC00}
EC^{**}&=C_0\times {\rm Pr}(N=0)+\sum_{n=1}^{\infty}p(N)\times {\rm Pr}(N=n),
\end{align}
where $p(N)$ is obtained in Lemma \ref{lem.c1}. For \eqref{eq.EC00}, with the PDF of $N$, we have the following equality chain:
\begin{align}\label{eq.proof_c1}
&EC^{**} = C_0e^{-\lambda\pi d^2}+\sum_{n=1}^{\infty} \big(\epsilon+\int_{0}^{\beta B}\big(1-\Phi(\frac{x}{\beta}+Q-B)\big)^n\big) \notag \\
&\quad\quad\quad\quad\quad\quad\quad\quad\quad\quad \times \frac{(\lambda\pi d^2)^n}{n!}\exp(-\lambda\pi d^2) dx \notag \\
& = C_0e^{-\lambda\pi d^2}+\epsilon(1-\exp(-\lambda\pi d^2))\notag \\
&~+\int_{0}^{\beta B}\sum_{n=1}^{\infty} (1-\Phi(\frac{x}{\beta}+Q-B))^n \frac{(\lambda\pi d^2)^n}{n!} \exp(-\lambda\pi d^2)dx \notag \\
& =\epsilon+(C_0-\epsilon)\exp(-\lambda\pi d^2) + \exp(-\lambda\pi d^2)\notag \\
& \quad\quad \quad \times \int_{0}^{\beta B}\big[\exp\big(\lambda\pi d^2(1-\Phi(\frac{x}{\beta}+Q-B))\big)-1\big]dx \notag \\
& =\epsilon+(C_0-\epsilon-\beta B)\exp(-\lambda\pi d^2) \notag \\
&\quad + \int_{0}^{\beta B} \exp\bigg(\frac{\lambda\pi d^2\big({\rm erfc}\big(\frac{x/\beta+Q-B-\mu}{\sqrt{2}\sigma}\big)-2\big)}{2}\bigg)dx,
\end{align}
where the second equality in \eqref{eq.proof_c1} follows from switching the summarization and the integration operations, and the last equality in \eqref{eq.proof_c1} holds from the fact $1-\Phi(x)={\rm erfc}(x/\sqrt{2})/2$.

\subsection{Proof of Theorem \ref{Theo1}}
First, consider the case of $EC_{\rm hom}(p)=\widetilde{EC}_{\rm hom}(p)$ for $p\in[p_{\rm th1}, p_{\rm th2}]$, where $p_{\rm th1}\triangleq \epsilon+\beta(B+\mu-Q-2\sqrt{2}\sigma)$ and $p_{\rm th2}\triangleq \epsilon+\min\{\beta(B+\mu-Q+2\sqrt{2}\sigma),\beta B\}$. To this end, we establish the convexity of $\widetilde{EC}_{\rm hom}(p)$ of $p$ under the condition of $B+\mu \leq Q$.
 \begin{lemma}\label{lem.new}
Suppose that $B+\mu \leq Q$. The function $\widetilde{EC}_{\rm hom}(p)$ is a convex function of $p\in [p_{\rm th1},\beta(B+\mu-Q+2\sqrt{2}\sigma)]$.
 \end{lemma}
 \begin{IEEEproof}
 This lemma can be verified by checking its positiveness of the second-order derivative of $\widetilde{EC}_{\rm hom}(p)$. At first, the first-order derivative of $\widetilde{EC}_{\rm hom}(p)$ is expressed as
 \begin{align*}
  & \widetilde{EC}^\prime_{\rm hom}(p)= 1 - \exp\big(-\lambda\pi d^2 \Omega(p)\big) \Bigg( 1 \notag \\
 & \quad\quad +\frac{\lambda\pi d^2(C_0-p)}{\sqrt{2\pi}\sigma \beta}\exp\Big(-\Big[\frac{\frac{p-\epsilon}{\beta}+Q-B-\mu}{\sqrt{2}\sigma}\Big]^2\Big)\Bigg).
 \end{align*}
 Based on $\widetilde{EC}^\prime_{\rm hom}(p)$, the second-order derivative of $\widetilde{EC}_{\rm hom}(p)$ is then given by $\widetilde{EC}^{\prime\prime}_{\rm hom}(p)=$
 \begin{align*}
  &\frac{\lambda\pi d^2}{\sqrt{2\pi}\sigma\beta}\exp(-\lambda\pi d^2\Omega(p)-\Big[\frac{\frac{p-\epsilon}{\beta}+Q-B-\mu}{\sqrt{2}\sigma}\Big]^2)\notag\\
& \times\Bigg(2 + (C_0-p)\big[\frac{\lambda\pi d^2}{\sqrt{2\pi}\sigma\beta}\exp\big(-\big[\frac{\frac{p-\epsilon}{\beta}+Q-B-\mu}{\sqrt{2}\sigma}\big]^2\big)\notag \\
 &+\frac{\frac{p-\epsilon}{\beta}+Q-B-\mu}{\sigma^2\beta}\big]\Bigg).
 \end{align*}
Under the condition of $B+\mu\leq Q$ in Lemma \ref{lem.new}, it is verified that
\begin{align*}
&2+(C_0-p)\Bigg[\frac{\lambda\pi d^2}{\sqrt{2\pi}\sigma\beta}\exp\Big(-\big[\frac{\frac{p-\epsilon}{\beta}+Q-B-\mu}{\sqrt{2}\sigma}\big]^2\Big)\notag \\
& \quad\quad\quad \quad\quad\quad +\frac{\frac{p-\epsilon}{\beta}+Q-B-\mu}{\sigma^2\beta}\Bigg]\geq 0,
 \end{align*}
and hence $\widetilde{EC}^{\prime\prime}_{\rm hom}(p)\geq0$ for $p\in [p_{\rm th1},\beta(B+\mu-Q+2\sqrt{2}\sigma)]$. By the non-negativeness of $\widetilde{EC}_{\rm hom}^{\prime\prime}(p)$, it follows that $\widetilde{EC}_{\rm hom}(p)$ is a convex function of $p$.
 \end{IEEEproof}

Next, based on the convexity of $\widetilde{EC}_{\rm hom}(p)$ in Lemma \ref{lem.new}, the unique minimum $\hat{p}$ of $\widetilde{EC}_{\rm hom}(p)$ can be obtained by solving the equation $\widetilde{EC}(\hat{p})=0$. In addition, it holds that
\begin{align}\label{eq.positive}
& \widetilde{EC}^\prime_{\rm hom}(p_{\rm th1}) \notag \\
&= -\frac{\lambda\pi d^2(C_0-\epsilon-\beta(B+\mu-Q-2\sqrt{2}\sigma))\exp(-4)}{\sqrt{2\pi}\sigma\beta}<0 \notag \\
& \widetilde{EC}^\prime_{\rm hom}(\epsilon+\beta(B+\mu-Q+2\sqrt{2}\sigma)) = 1-\exp(-\lambda\pi d^2)\Big(1+\notag \\
&\quad\quad \frac{\lambda\pi d^2(C_0-\epsilon-\beta(B+\mu-Q+2\sqrt{2}\sigma))\exp(-4)}{\sqrt{2\pi}\sigma\beta}\Big) >0 \notag \\
&\widetilde{EC}^\prime_{\rm hom}(C_0)=1-\exp(-\lambda\pi d^2)>0.
\end{align}
From \eqref{eq.positive}, it follows that the global minimum $\hat{p}\in[p_{\rm th1}, \min\{\epsilon+\beta(B+\mu-Q+2\sqrt{2}\sigma),C_0\}]$.
Besides, the value of $\widetilde{EC}^\prime_{\rm hom}(\epsilon)=1-\exp(-\lambda\pi d^2\Omega(\epsilon))\big( 1 +\frac{\lambda\pi d^2(C_0-\epsilon)}{\sqrt{2\pi}\sigma \beta}\exp\big(-\big[\frac{Q-B-\mu}{\sqrt{2}\sigma}\big]^2\big) \big)$ can either be negative or nonnegative. More specifically, if $\widetilde{EC}^\prime_{\rm hom}(\epsilon)\geq 0$, then $\hat{p}\leq \epsilon$. On the other hand, if $\widetilde{EC}^\prime_{\rm hom}(\epsilon)\leq 0$, then $\hat{p}\geq \epsilon$.

On the other hand, if $B+\mu>Q$, it is not guaranteed that $\widetilde{EC}^{\prime\prime}_{\rm hom}(p)\geq 0$ for any $p\in[p_{\rm th1},\beta(B+\mu-Q+2\sqrt{2}\sigma)]$. It thus implies that there may exist multiple candidate solutions to $\widetilde{EC}^{\prime}_{\rm hom}(p)= 0$ for $p\in[p_{\rm th1},\beta(B+\mu-Q+2\sqrt{2}\sigma)]$. In this case, one can compute the corresponding expected cost under different prices, and then select the best solution $\hat{p}$ that returns the minimal expected cost value.

Last, consider the cases of $p-\epsilon\geq p_{\rm th}$. Traveler T's expected cost $EC_{\hom}(p)=C_0+(p-C_0)(1-\exp(-\lambda\pi d^2))$ is an increasing linear function of $p$, for which $EC_{\hom}(C_0)=C_0$. In addition, $EC_{\hom}(p)> \widetilde{EC}_{\rm hom}(p_{\rm th})$ for any $p \geq p_{\rm th}$.

To summarize, the solution to minimize ${EC}_{\rm hom}(p)$ in the range $[\epsilon,C_0]$ is $\max\{\epsilon, \hat{p}\}$ and Theorem~\ref{Theo1} follows.

\subsection{Proof of Proposition \ref{prop.variance}}
Define a function $F(\sigma)=\widetilde{EC}_{\rm hom}(p)$ with respect to $\sigma>0$, i.e.,
\begin{align}
F(\sigma)=C_0+(p-C_0)\Big(1-\exp(-\lambda\pi d^2\Omega(p))\Big).
\end{align}
The first-order derivative $F^\prime(\sigma)=\frac{\partial F(\sigma)}{\partial \sigma}$ is then
\begin{align}\label{eq.F}
F^\prime(\sigma) = &-\sigma^2(p-C_0) \frac{\lambda\pi d^2(p-\epsilon+\beta(Q-B-\mu))}{\sqrt{2\pi}\beta}\notag\\
&\quad\quad\quad\quad\quad \times\exp\Big(-\lambda\pi d^2\Omega(p)-\Upsilon^2\Big),
\end{align}
where $\Upsilon \triangleq \frac{p-\epsilon+\beta(Q-B-\mu)}{\sqrt{2}\sigma\beta}$. If $B+\mu\leq Q$, we can show that $F^\prime(\sigma)>0$ for any $p\in[\epsilon, C_0]$ by checking each multiplication factor in \eqref{eq.F}. Hence, the traveler's expected cost $EC_{\rm hom}(p)$ increases with the data usage variance. Proposition 4.1 thus follows.

\subsection{Proof of Theorem \ref{Theo2}}
For $k=1,\ldots,K$, we define the following function
\begin{align}
g_k(p) \triangleq C_0+(p-C_0)\big(1-\exp(-\sum_{j=1}^k \lambda_j\pi d^2 \Omega_j(p))\big),
\end{align}
where $\epsilon\leq p\leq C_0$. We first establish the convexity of $g_k(p)$ under the condition $B+\mu_k\leq Q_k$, $k\in{\cal K}$. in the following lemma.
\begin{lemma}\label{lem.convex_k}
Suppose that $B+\mu_k\leq Q_k$, $k\in{\cal K}$. The function $g_k(p)$ for any $k\in{\cal K}$ is a convex function of $p\in[\epsilon,C_0]$.
\end{lemma}
\begin{IEEEproof}
This lemma can be verified by checking the positiveness of the second-order derivative of $g_k(p)$ with respect to $p$. The first-order derivative of $g_k(p)$ with $p$ is given by
\begin{align*}
& g_k^{\prime}(p) = 1-\exp\Big(-\sum_{j=1}^k\lambda_j\pi d^2\Omega_j(p)\Big)\Bigg[ 1 \notag\\
& \quad\quad + \sum_{j=1}^k\frac{\lambda_j\pi d^2(C_0-p)\exp(-(\frac{p-\epsilon+\beta_j(Q_j-B-\mu_j)}{\sqrt{2}\sigma_j\beta_j})^2)}{\sqrt{2\pi}\sigma_j\beta_j}\Bigg],
\end{align*}
and the second-order derivative of $g_k(p)$ is
\begin{align*}
&g_k^{\prime\prime}(p) = \sum_{j=1}^K\frac{\lambda_j\pi d^2}{\sqrt{2\pi}\sigma_j\beta_j}\exp\Bigg(-\sum_{j=1}^k\lambda_j\pi d^2\Omega_j(p)\notag\\
&\quad-\Big[\frac{p-\epsilon+\beta_j(Q_j-B-\mu_j)}{2\sigma_j^2\beta_j^2}\Big]^2\Bigg)\Bigg[2+(C_0-p)\notag\\
& \quad \times \sum_{j=1}^K \Bigg(\frac{\lambda_j\pi d^2}{\sqrt{2\pi}\sigma_j\beta_j}\exp\Big(-\Big[\frac{p-\epsilon+\beta_j(Q_j-B-\mu_j)}{2\sigma_j^2\beta_j^2}\Big]^2\Big)\notag \\
&\quad +\frac{p-\epsilon+\beta_j(Q_j-B-\mu_j)}{2\sigma_k^2\beta_j^2}\Bigg)\Bigg].
\end{align*}
Under the condition of $B+\mu_j\leq Q_j$, it follows that $g_k^{\prime\prime}(p)\geq 0$ for any $k\in{K}$. As a result, the function $g_k(p)$, $k\in{\cal K}$, is a convex function of $p\in[\epsilon,C_0]$, and thus lemma \ref{lem.convex_k} follows.
\end{IEEEproof}

Denote by $p^*_k$ the optimal solution to $g_k^\prime(p_k)=0$. Since that $g^\prime_k(\epsilon)<0$ and $g^{\prime}_k(C_0)>0$, $k\in{\cal K}$, it holds that $\epsilon\leq p^*_k\leq C_0$, $k\in{\cal K}$.

On the other hand, if $B+\mu_k> Q_k$, $k\in{\cal K}$, the convexity of $g_k(p)$ is not guaranteed for $p\in[\epsilon,C_0]$. This implies that there may exist multiple solutions to $g_k^\prime(p)=0$ for $p\in[\epsilon,C_0]$. In this case, one can compute the corresponding expected costs under different price values via a one-dimensional exhaustive search, and then select the best solution $p_k^*$ that returns the minimum expected cost for traveler T.

Based on the above analysis, consider the case of $C_0\leq T_1$. Since the pH data sharing probability $\Omega_k(p)=0$, $\forall k\in{\cal K}$, traveler T has to pay roaming fee $C_0$ to the WNO.

Next consider the case of $T_k\leq C_0\leq T_{k+1}$ for $k=1,\ldots,K-1$. Due to the fact that $EC_{\rm het}(p)=g_k(p)$ and its convexity of $g_k(p)$ in the range $[\epsilon,C_0]$, the unique solution $p^*_k$ to minimize $EC_{\rm het}(p)$ satisfies $\epsilon\leq p_k^* \leq C_0 \leq T_{k+1}$ in this case.

For the case of $T\geq T_K$, we have $EC_{\rm het}(p)=g_K(p)$. Based on the convexity of $g_K(p)$ and $g_K^\prime(p_K^*)=0$, it is clear that $p_K^*$ is the optimal price for traveler T to minimize $g_K(p)$.

\subsection{Derivations of $\mathbb{P}^{\tt UB1}_{\tt mul}(p)$ and $\mathbb{P}_{\tt mul}^{\tt UB2}(p)$}
We present the detailed calculation for obtaining the upper bound expressions in \eqref{eq.P_UB1} and \eqref{eq.P_UB2} as follows.
\begin{align}\label{eq.upper1}
&\mathbb{P}_{\tt mul}^{\tt UB1}(p) = \sum_{M=0}^{\infty}\sum_{N=1}^{\infty}\sum_{N_y=1}^{N}{N\choose N_y}\big(\Omega(p)\big)^{N_y}\big(1-\Omega(p)\big)^{N-N_y}\notag\\
&\quad\quad \quad\quad \times\frac{(\lambda\pi d^2)^N \exp(-\lambda\pi d^2)}{N!}\frac{(\lambda_t\pi d^2)^M\exp(-\lambda_t\pi d^2)}{M!}\notag \\
&=\sum_{M=0}^{\infty}\sum_{N=1}^{\infty}\big(1-(1-\Omega(p))^N\big)\frac{(\lambda\pi d^2)^N \exp(-\lambda\pi d^2)}{N!} \notag\\
&\quad\quad \quad\quad\quad \quad\quad\quad \quad \times \frac{(\lambda_t\pi d^2)^M\exp(-\lambda_t\pi d^2)}{M!}\notag\\
&=\Big(\sum_{N=1}^{\infty}\frac{(\lambda\pi d^2)^N}{N!}-\sum_{N=1}^{\infty}\frac{((1-\Omega(p))\lambda \pi d^2)^N}{N!}\Big)\notag\\
&\quad\quad \quad \times \exp(-\lambda\pi d^2)\Big(\sum_{M=0}^{\infty}\frac{(\lambda_t\pi d^2)^M}{M!}\Big)\exp(-\lambda_t\pi d^2) \notag \\
&=\Big(\exp(\lambda\pi d^2)-\exp((1-\Omega(p))\lambda\pi d^2)\Big)\exp(-\lambda\pi d^2)\notag \\
&=1-\exp(((1-\Omega(p))-1)\lambda\pi d^2),
\end{align}
where the fourth equality of \eqref{eq.upper1} follows from the fact that $\sum_{N=1}^{\infty}{(\lambda\pi d^2)^N}/{N!}=\exp(\lambda\pi d^2)-1$, $\sum_{N=1}^{\infty}{((1-\Omega(p))\lambda \pi d^2)^N}/{N!}=\exp((1-\Omega(p))\lambda \pi d^2)-1$, and $\sum_{M=0}^{\infty}(\lambda_t\pi d^2)^M/{M!}=\exp(\lambda_t\pi d^2)$.

Likewise, for $\mathbb{P}^{\tt UB2}_{\tt mul}(p)$ in \eqref{eq.P_UB2}, we have
\begin{align}\label{eq.upper2}
&\mathbb{P}^{\tt UB2}_{\tt mul}(p)=\notag \\
& \sum_{M=0}^{\infty}\sum_{N=1}^{\infty}\frac{\sum_{N_y=1}^{N} N_y{N\choose N_y}(\Omega(p))^{N_y}(1-\Omega(p))^{N-N_y}}{M+1}\notag \\
&\quad\quad\quad \times \frac{(\lambda\pi d^2)^N \exp(-\lambda\pi d^2)}{N!}\frac{(\lambda_t\pi d^2)^M\exp(-\lambda_t\pi d^2)}{M!}\notag \\
&=\Bigg(\sum_{N=1}^{\infty}\Omega(p)N\frac{(\lambda\pi d^2)^N\exp(-\lambda\pi d^2)}{N!}\Bigg) \notag\\
&\quad\quad\quad\quad\quad\quad \times \Bigg(\sum_{M=0}^{\infty}\frac{(\lambda_t\pi d^2)^M\exp(-\lambda_t\pi d^2)}{(M+1)!}\Bigg) \notag \\
&=\lambda q\Omega(p)\pi d^2 \exp(-\lambda\pi d^2) \Bigg(\sum_{N=1}^{\infty}\frac{(\lambda\pi d^2)^{N-1}}{(N-1)!}\Bigg)\notag \\
&\quad\quad\quad \quad\quad\quad \times \Bigg(\frac{\exp(-\lambda_t\pi d^2)}{\lambda_t\pi d^2}\sum_{M=0}^{\infty}\frac{(\lambda_t\pi d^2)^{M+1}}{(M+1)!}\Bigg)
\notag \\
&=\lambda \Omega(p)\pi d^2 \exp(-\lambda\pi d^2) \exp(\lambda\pi d^2)\frac{\exp(-\lambda_t\pi d^2)}{\lambda_t\pi d^2}\notag\\
&\quad\quad\quad\quad\quad\quad \times \big(\exp(\lambda_t\pi d^2)-1\big)
\notag \\
&= \frac{\lambda\Omega(p)\big(1-\exp(-\lambda_t\pi d^2)\big)}{\lambda_t},
\end{align}
where the second equation of \eqref{eq.upper2} follows from the fact that $\sum_{n=1}^N n{N \choose n}x^n(1-x)^{N-n}=Nx$ for $0<x<1$, and the fourth equation of \eqref{eq.upper2} holds from the fact that $\sum_{M=0}^{\infty}(\lambda_t\pi d^2)^{M+1}/{(M+1)!}=\exp(\lambda_t\pi d^2)-1$ and $\sum_{N=1}^{\infty}{(\lambda\pi d^2)^{N-1}}/{(N-1)!}=\exp(\lambda\pi d^2)$.

\subsection{Proof of lemma \ref{lem.mediate}}
First, we prove the function $f(p)$ has a unique solution $\tilde{p}$, where $p\in[\epsilon,C_0]$, and $f(p)$ is defined as
\begin{align}
f(p)&\triangleq  \mathbb{P}_{\rm mul}^{\rm UB2}(p)\times p + (1-\mathbb{P}_{\rm mul}^{\rm UB2}(p))\times C_0\notag \\
& =C_0+\frac{\lambda \Omega(p)(1-\exp(-\lambda_t\pi d^2))}{\lambda_t}(p-C_0).
\end{align}
The first-order derivative of $f(p)$ with respect to $p$ is given as
\begin{align*}\label{eq.EC1_prime}
& f^{\prime}(p) =\frac{\lambda(1-\exp(-\lambda_t\pi d^2))}{\lambda_t}\Big(\Omega(p) \notag\\
&\quad +\frac{1}{\sqrt{2\pi}\sigma\beta}\exp\big(-\big[\frac{p-\epsilon+\beta(Q-B-\mu)}{\sqrt{2}\sigma\beta}\big]^2\big)(p-C_0)\Big).
\end{align*}

Suppose that $f^\prime(\tilde{p})=0$. Let $p=\tilde{p}-\Delta$ where $0<\Delta<\tilde{p}-\epsilon$. Then $\epsilon<p<\tilde{p}$. The first-order derivative of $\exp\big(-\left[\frac{p-\epsilon+\beta(Q-B-\mu)}{\sqrt{2}\sigma\beta}\right]^2\big)(p-C_0)$ of $p\in[\epsilon,C_0]$ is positive, i.e.,
\begin{align}
&\exp\Bigg(-\Big[\frac{p-\epsilon+\beta(Q-B-\mu)}{\sqrt{2}\sigma\beta}\Big]^2\Bigg) \notag \\
& \quad\quad\quad +(C_0-p)\frac{p-\epsilon+\beta(Q-B-\mu)}{\sigma^2\beta^2}>0,
\end{align}
it follows that $\exp\big(-\big[\frac{p-\epsilon+\beta(Q-B-\mu)}{\sqrt{2}\sigma\beta}\big]^2\big)(p-C_0)<\exp\big(-\big[\frac{\tilde{p}_1-\epsilon+\beta(Q-B-\mu)}{\sqrt{2}\sigma\beta}\big]^2\big)(\tilde{p}_1-C_0)$. Furthermore, together with the monotonically increasing $t(p)$ of $p$, we have $f^{\prime}(p)<f^{\prime}(\tilde{p})$. Likewise, letting $p=\tilde{p}+\Delta$ where $0<\Delta<C_0-\tilde{p}$, we have $f^\prime(p)>f^{\prime}(\tilde{p})=0$.

Based on above analysis for $f^{\prime}(p)$ of $p\in[\epsilon,C_0]$, it is verified that $\tilde{p}$ is the unique solution to minimize $f(p)$. Then, by checking $f^{\prime}(\Omega^{-1}(\lambda_t/\lambda))<0$, it follows that $\tilde{p}>\Omega^{-1}(\lambda_t/\lambda)$. Taking into account the constraint of $\epsilon\leq p \leq \Omega^{-1}(\lambda_t/\lambda)$, the optimal solution is thus $p_{\rm mul}^*=\Omega^{-1}(\lambda_t/\lambda)$.


%

\begin{IEEEbiography}
{Feng Wang}(M'16) received the B.Eng. degree from Nanjing University of Posts and Telecommunications, China, in 2009, and the M.Sc. and Ph.D. degrees, both from Fudan University, China, in 2012 and 2016, respectively. He is currently an Assistant Professor with the School of Information Engineering, Guangdong University of Technology, China. From 2012 to 2013, he was a Research Fellow with the Department of Communication Technology, Sharp Laboratories of China. From Jan. 2017 to Sep. 2017, he was a Post-Doctoral Research Fellow with the Engineering Systems and Design Pillar, Singapore University of Technology and Design. His research interests include signal processing for communications, energy harvesting wireless communications, and mobile edge computing.
\end{IEEEbiography}


\begin{IEEEbiography}
{Lingjie Duan}(S'09-M'12-SM'16) received the Ph.D. degree from The Chinese University of Hong Kong in 2012. In 2011, he was a Visiting Scholar with the University of California at Berkeley, CA, USA. He is currently an Assistant Professor with the Engineering Systems and Design Pillar, Singapore University of Technology and Design, Singapore. His current research interests include network economics and game theory, cognitive and cooperative communications, energy harvesting wireless communications, and mobile crowdsourcing. He is an Editor of the IEEE TRANSACTIONS ON WIRELESS COMMUNICATIONS and IEEE COMMUNICATIONS SURVEYS AND TUTORIALS. In 2016, he was a Guest Editor of the IEEE JOURNAL ON SELECTED AREAS IN COMMUNICATIONS Special Issue on Human-in-the-loop Mobile Networks, and was also a Guest Editor of the IEEE Wireless Communications Magazine for feature topic Sustainable Green Networking and Computing in 5G Systems. He served as the Program Co-Chair at the IEEE ICC 2019 Cognitive Radio and Networks Symposium, the VTC-2017 Future Trends and Emerging Technologies track, the IEEE INFOCOM 2014 GCCCN Workshop, the Wireless Communication Systems Symposium of the IEEE ICCC 2015, the GCNC Symposium of the IEEE ICNC 2016, and the IEEE INFOCOM 2016 GSNC Workshop. He has also served as a Technical Program Committee (TPC) Member of many leading conferences in communications and networking (e.g., ACM MobiHoc, IEEE INFOCOM, SECON, and WiOPT). He was a recipient of the 2016 SUTD Excellence in Research Award, the 10th IEEE ComSoc Asia-Pacific Outstanding Young Researcher Award in 2015, and the Hong Kong Young Scientist Award (Finalist in Engineering Science track) in 2014.
\end{IEEEbiography}

\begin{IEEEbiography}
{Jianwei Niu}(SM'12) received the M.S. and Ph.D. degrees in computer science from Beihang University, Beijing, China, in 1998 and 2002, respectively. He was a Visiting Scholar with the School of Computer Science, Carnegie Mellon University, Pittsburgh, PA, USA, from 2010 to 2011. He is currently a Professor with the School of Computer Science and Engineering, Beihang University. He has authored or coauthored more than 150 referred papers and has filed more than 30 patents. His current research interests include wireless sensor networks and mobile and pervasive computing. He served as a TPC member for InfoCom, PerCom, ICC, WCNC, GLOBECOM, and LCN. He was a recipient of the New Century Excellent Researcher Award from the Ministry of Education of China in 2009, the first prize of the Technical Invention Award from the Ministry of Education of China in 2012, the Innovation Award from the Nokia Research Center, and the Best Paper Award at JNCA 2015, IEEE ICC 2013, WCNC 2013, ICACT 2013, CWSN 2012, and GreenCom 2010. He served as the DySON Workshop Co-Chair for INFOCOM 2014, the Program Chair for the IEEE SEC 2008, and the Executive Co-Chair for the TPC of CPSCom 2013. He has served as an Associate Editor for the International Journal of Ad Hoc and Ubiquitous Computing, Journal of Network and Computer Applications, and Mobile Networks and Applications.
\end{IEEEbiography}

\end{document}